\documentclass{cargese}

\let\footnote\savefootnote
\let\footnotetext\savefootnotetext 
\let\footnoterule\savefootnoterule 
\normallatexbib
\def \ci{\cite}

\newcommand{\p}[1]{(\ref{#1})}
\newcommand{\bt}[1]{{\bar t}}

\def \inti { \int^{2\pi}_0 {d\s \ov 2\pi} }

\def \rH {{\rm H}} \def \rE {{\rm E}} 

\def \sql {{\sqrt{\l}}\ }
\def \del{\partial}
\def \a {\alpha}

\def\g{\gamma}
\def\s{\sigma}

\def\ov{\over}
\def\la{\label}

\def\LL{{\cal J }}

\def\b{\beta}
\def\l{\lambda}

\def \adss{$AdS_5 \times S^5$\ }

\def \sql {\sqrt{\lambda} }

\def \p {\phi}

\def \ov {\over}
\def \s{\sigma}

\def \ta{\tau}

\def \ha {{1 \over 2}}

\def \k {\kappa}
\def\foot{\footnote}

\def \L {\Lambda}

\newcommand{\rf}[1]{(\ref{#1})}

\def \four{{\textstyle {1\ov 4}}}
 
\def\det{\hbox{det}}
\def\be{\begin{equation}}
\def\ee{\end{equation}}

\def \ci {\cite}

\def \foot {\footnote}
\def \bi{\bibitem}

\def \tr {{\rm tr}}
\def \ha {{1 \over 2}}
\def \td {\tilde}

\def \ci{\cite}
\def \N {{\cal N}}

\def\S{{\cal S} }

\def \XX {{\rm X}}

\def \n {\nu}

\def \del{\partial}
\def \Tr {{\rm Tr}}

\def \tr {{\rm tr}}
\def \ha {{1 \over 2}}
\def \ov {\over}

\def \sql {{\sqrt{\l}}\ }

\def \ta {\tau}

\def \D {{\Delta}}
\def \ta {\tau}

\def \sql {{\sqrt{\l}}\ }
\def \del{\partial}

\def\b{\beta}
\def\l{\lambda}

\def \ov {\over}
\def \s{\sigma}

\def \ta{\tau}

\def \ha {{1 \over 2}}

\def \k {\kappa}

\def \four{{\textstyle {1\ov 4}}}
 
\def\det{\hbox{det}}

\def \foot {\footnote}
\def \bi{\bibitem}

\def \tr {{\rm tr}}
\def \ha {{1 \over 2}}
\def \td {\tilde}
\def \ci{\cite}
\def \N {{\mathcal N}}

\def\S{{\mathcal S} }

\def \XX {{\rm X}}

\def \d {\del}

\def \L {\Lambda}

\def \tr {{\rm tr\ }}

\def\a{\alpha}
\def\b{\beta}

\def\p{\phi}

\def\P{{\bf P}}

\def\Tr{{\rm  Tr}}
\def\tr{{\rm  tr}}

\def \del{\partial}
\def \a {\alpha}

\def\g{\gamma}
\def\s{\sigma}

\def\ov{\over}

\def\LL{{\mathcal L }}

\def\b{\beta}
\def\l{\lambda}

\def \De {{\mathcal D}}

\def \k {\kappa}
\def\foot{\footnote}
\def \four{{\textstyle {1\ov 4}}}
 
\def\det{\hbox{det}}
\def \ci {\cite}

\def \foot {\footnote}
\def \bi{\bibitem}

\def \tr {{\rm tr}}
\def \ha {{1 \over 2}}
 
\def \vv {{\rm v}}  
\def \tl {{\tilde \l}}
\def \XX {{\rm X}}
\def \ta {{\tilde \a}}
\def \fo { {1\ov 4}}
\def \ep {\epsilon}
\def \inti {{\int^{2\pi}_0 {d \sigma \ov 2 \pi}}}

\def \d {\partial}

\def \el {\ell}
\def \Tr {{\rm Tr}}
\def \P {\Phi}
\def \l  {\lambda}
\def \tl {{\tilde \l}}
\def \bl {{\tilde \l}}

\def \bv {v^*} 
\def \vv {{\rm v}}
\def \LL {{\mathcal L}}

\def \N {{\mathcal N}} 
\def \S {{\rm S}} 
\def \vn {\vec n}
\def \tl {\td \l} 
\def \td {\tilde} 
\def \Prod {\Pi}
\def \O {{\mathcal O}}

\def \D {\Delta}
\def \N {{\mathcal N}}

\def \m {\mu}
\def \vs {\vec \s} 
\def \ie {i.e.}

\def \cD {{\cal D}}

\def  \le  {\l_{\rm eff}} 

\def \rS {{\rm S}}
\newcommand{\bra}[1]{\mbox{$\langle #1 |$}}
\newcommand{\ket}[1]{\mbox{$| #1 \rangle$}}

\begin{document}
%------------ article title  ------------------->>
% For a long title use \\ to cut lines.
% In that case, supply  alternate version of the title
% in square brackets, (it will go in the Table of contents during final
% production of the book.
% \articletitle[Short title]{The long version \\ of this title}

\articletitle{Semiclassical strings  and   AdS/CFT}

%% optional, to supply a shorter version of the title for the running head:
%%\chaptitlerunninghead{}

\author{A.A. Tseytlin\foot{Also at Imperial College, London and
Lebedev Physics Institute, Moscow.  }}

%% multiple authors at the same institution may be separated with \\
%% like in \author{Samuel Bostaph\\
%%                 and Gregor Kariotis}

%% Your Institution and address. May cut into  separated lines with \\

\affil{Physics Department, The Ohio State
University,\\
 Columbus OH43210-1106, USA  }    

% optional email address
\email{tseytlin.1@osu.edu}

%% Repeat the above for multiple authors at different institutions.
%% \author{ }
%% \affil{ }
%% \email{ }

% optional abstract
\begin{abstract}
We discuss AdS/CFT duality in the sector of  ``semiclassical''
string  states with large  quantum numbers.
We review the coherent-state effective action 
approach, in which similar  2d sigma model  actions 
appear  from  the  $AdS_5 \times  S^5$
 string action and  from the integrable 
spin chain Hamiltonian representing the N=4 super Yang-Mills dilatation operator.  
We consider mostly  the leading-order  terms in the 
energies/anomalous dimensions which match but comment also on
  higher-order corrections. 
 \end{abstract}

%------------ body of article ------------------->>
% Write your article here. 
% Note that the \section{section title}
% command allows for the form \section[short title]{very long\\ title}
% Idem for \subsection and \subsubsection
%------------ end of article ------------------->>

%%%%%%%%%%%%%%%%%%%%%%%%%%%%%%%%%%%%%%%%%%%%
\setcounter{equation}{0}
%%%%%%%%%%%%%%%%%%%%%%%%%%%%%%%%%%%%%%%%%%%%
\setcounter{footnote}{0}

%%%%%%%%%%%%%%%%%%%%%%%%%%%%%%%%%%%%%%%%%%%%%%%%%%%%%%%%%%
\section{Introduction}
%%%%%%%%%%%%%%%%%%%%%%%%%%%%%%%%%%%%%%%%%%%%%%%%%%%%%%%%%%%%%%%

The    $\N=4$ SYM theory with $SU(N)$ gauge group 
 is  a family of conformal theories 
 parametrized by the  two  numbers -- $N$ and $g_{_{\rm YM}}$. 
Four-dimensional conformal theories  have apparently much less
symmetry than their two-dimensional cousins  and thus 
should be much harder to solve (i.e. to determine their 
spectrum of  dimensions of conformal primary operators 
and their correlation functions). There are strong indications 
that this problem may simplify in the planar  ($N\to \infty $, 
\ $\lambda\equiv g_{_{\rm YM}}^2N$=fixed)  limit. 
In the large $N$ limit there are no formal objections to
integrability of a 4-d quantum field theory, and, indeed, 
 the AdS/CFT duality  \ci{malda,gub,wit} 
implies  the existence of a  hidden integrable 
2-d structure corresponding to that of \adss string sigma model
(on a sphere or a cylinder). 
A  major first  
step towards the solution of the  SYM  theory  would be to 
determine the spectrum of anomalous dimensions  $\Delta (\l)$ 
of the  primary operators 
built out of  products of local gauge-covariant fields.

The AdS/CFT duality implies the equality between the $AdS_5$ 
energies $E$ of quantum  closed string states (as functions of
 the effective string 
tension $T= {R^2  \ov 2 \pi\a'} = {\sql \ov 2 \pi}$   and quantum numbers like 
$S^5$ angular momenta  $J_i$) 
and the dimensions  $\Delta$ of the corresponding 
local SYM operators (see, e.g., \ci{oog}).  
%One may hope to try to  uncover  this  duality   in certain special 
%limits.
%  and in the process  get an insight into the structure of 
%functions that determine the anomalous dimensions to all orders in
%$\l$. 
%On more detailed level, one 
To check the duality  one  would like also  to understand  
 how strings
 ``emerge'' from the field theory, 
 in particular,  which (local,  single-trace) 
 gauge theory operators \ci{POL} 
 correspond to which ``excited'' string states and 
 how one may   verify
the matching of  their dimensions/energies 
beyond the well-understood BPS/supergravity sector.  
One  would  then like  to use the duality as a guide  to deeper  
level of the structure of quantum SYM theory.
 For example, the 
 results motivated by comparison to string theory 
may allow one to ``guess'' the general structure  of the SYM 
anomalous dimension matrix
 and may also  suggest new methods of computing anomalous
 dimensions  in less supersymmetric gauge theories.

Below we shall review some  recent progress  in  checking AdS/CFT
correspondence in a subsector of non-BPS  string/SYM  states
with large quantum numbers.

%%%%%%%%%%%%%%%%%%%%%%%%%%%%%%%%%%%%%%%%%%%%%%%%%%%%%%%%%%%
\subsection {Generalities} 
%%%%%%%%%%%%%%%%%%%%%%%%%%%%%%%%%%%%%%%%%%%%%%%%%%%%%%%%%%%%%%%%%

%%%%%%%%%%%%%%%%%%%%%%%%%%%%%%%%%%%%%%%%%%%%%%%%%%%%%%%%%%%%%%%%%
Let us start with brief remarks on the SYM and the 
string sides of the duality.
The  SYM  theory contains a gauge field, 6 scalars $\p_m$ 
and 4 Weyl fermions, all in adjoint representation of $SU(N)$. 
It has global conformal and R-symmetry, i.e. is invariant  under
 $SO(2,4) \times SO(6)$. To determine (in the planar limit) scaling 
 the 
dimensions  of local 
gauge-invariant operators one,  in general,  needs to find the anomalous 
dimension matrix to all orders in $\l$ and then to diagonalize it. 
The special case is that of chiral primary or BPS operators (and their descendants)
$\tr ( \p_{\{ m_1 } ... \p_{ m_k\} } )$ whose dimensions are protected, i.e. 
do not depend on $\l$. The problem of finding dimensions 
appears to simplify also in the  case of ``long'' operators 
containing  large number of fields  under the trace. One example 
is provided by ``near-BPS'' operators \ci{bmn} 
  like $\tr ( \P_1^J \P_2^n ...)+ ...$, 
where  $J \gg n$, and $\P_k = \p_k + i \p_{k+3}$,  \ $k=1,2,3$. 
Below we shall  consider ``far-from-BPS'' operators 
like  $\tr ( \P_1^{J_1} \P_2^{J_2} ...)+ ...$,  with 
$J_1 \sim J_2 \gg 1$.

The type IIB string action in \adss space  
has the following structure 
\be \la{ss} 
I = -\ha T \int  d \tau \int^{2 \pi}_0  d \s \left(
\del^p Y^\m \del_p Y^\nu \eta_{\m\n} + 
\del^p X^m \del_p X^n \delta_{mn}  + ... \right)
\ , \ee 
where $ Y^\m Y^\nu \eta_{\m\n} =-1, \  \ X^m X^n \delta_{mn}=1 \ , \ 
\eta_{\m\n} = (-++++-)$,  
$T= { \sql \ov 2 \pi} $ and dots stand for the fermionic  terms 
\ci{mt}  that ensure that this model defines a 2-d  conformal field theory.
The closed string states can  be  classified by the values of the Cartan 
charges of the obvious symmetry group $SO(2,4) \times SO(6)$  
--  $E,S_1,S_2; J_1,J_2,J_3$, i.e. by the $AdS_5$ energy, two spins in $AdS_5$ and 3 spins in $S^5$.
 The mass shell (conformal gauge constraint)  condition then  gives a relation 
$E=E(Q, T)$. Here  $T$ is the string tension and 
$Q=(S_1,S_2,J_1,J_2,J_3; n_k)$ where $n_k$ stand for higher conserved charges
 (analogs of oscillation numbers in flat space). 
The BPS (chiral primary) string states are point-like (supergravity modes), 
near-BPS (BMN) states are nearly pointlike, while 
generic semiclassical far-from-BPS states are 
represented by extended closed string 
configurations. 

According to the AdS/CFT duality  quantum closed string
 states in \adss 
 should be dual to quantum SYM states at the boundary, i.e. in 
  $R \times S^3$ 
or, via radial quantization, 
 to local single-trace operators at the origin of $R^4$. 
 Such operators have the following structure 
 
 \noindent
 $ \tr (  D_{1+ i2}^{S_1} D_{3 + i 4 }^{S_2}
 \P_1^{J_1} \P_2^{J_2} \P_3^{J_2}...)+ ...$
 (where scalars and covariant  derivatives may be also 
 replaced by gauge field strength factors and fermions).
The energy of a string state  should then be equal to the dimension 
of the corresponding SYM operator, 
$E( Q,T)= \D ( Q, \l)$,  where on the SYM side  the charges $Q$ 
should characterise the eigen-operator of the anomalous dimension matrix. 
By analogy with the 
flat space   case  and ignoring $\a'$ corrections 
(i.e. assuming $R \to \infty$ or $\a' \to 0$) 
the excited  string states are expected to have energies
$E \sim m \sim { 1 \ov \sqrt{\a'}} \sim  \l^{1/4}$  \ci{gub}. 
This  represents 
a non-trivial prediction for strong-coupling asymptotics of SYM
 dimensions. The asymptotics may, however, be different 
in the limits where the charges $Q$ are also large,  
e.g., of order $\l^{1/2}$ as in the semiclassical limit 
\ci{gkp}.

In general, the natural (inverse-tension)  perturbative
expansion on the string side 
will be given by $E=\sum^\infty_{n=-1}  {c_n \ov (\sql)^n} $, while 
on the SYM side the usual planar perturbation theory  will give 
the eigenvalues of the anomalous dimension matrix as 
$\D= \sum^\infty_{n=0} a_n \l^n$. The AdS/CFT duality  implies 
that the two expansions should  be the 
strong-coupling and the weak-coupling asymptotics of the same function. 
To check the relation $E= \D$ is then a non-trivial  problem, except  
% This  can be done,  on the symmetry grounds, 
 in the case of 1/2  BPS (single-trace chiral primary) operators which are 
dual to the supergravity states  when 
%(``massless'' or ground state string modes): 
%in the BPS case
 the  energies/dimensions are protected from corrections
\ci{oog} and thus can be matched on the symmetry grounds.

%%%%%%%%%%%%%%%%%%%%%%%%%%%%%%%%%%%%%%%%%%%%%%%%%%%%%%%%%%%
\subsection {Semiclassical string states:  BMN  and  beyond } 
%%%%%%%%%%%%%%%%%%%%%%%%%%%%%%%%%%%%%%%%%%%%%%%%%%%%%%%%%%%%%%%%%

For generic non-BPS states  the situation with checking the duality
looked hopeless until 
the  remarkable suggestion of \ci{bmn}\foot{The history of the BMN limit 
is  somewhat non-trivial. It started with the 
important  observation that  the 
Penrose limit of the \adss space  \ci{papa}
leads to a maximally supersymmetric plane wave geometry supported by the 
Ramond-Ramond 5-form flux. 
Remarkably, the  
  \adss string theory  action \ci{mt} in this limit
  (i.e. Green-Schwarz string action in plane-wave background) 
   becomes 
essentially quadratic  and thus its spectrum can be found 
explicitly \ci{mets}. Motivated by that, ref. \ci{bmn} gave a 
dual SYM interpretation of the corresponding string states, 
suggested that string energies can be compared to perturbative 
SYM dimensions computed in the same limit,   and indeed 
directly checked this to the first  non-trivial order. 
Ref. \ci{gkp} then explained  that, on the string-theory side, 
 the BMN limit is nothing but a 
semiclassical expansion near a particular (point-like) string solution. 
 This interpretation 
 was further clarified and extended in \ci{ft1,tse1}, 
suggesting,  in particular,  how  one can in principle 
 compute corrections to the BMN limit (which was later done 
 in  full  detail in \ci{par} and especially in \ci{cal}).
This made it clear that the plane-wave connection  is not fundamental
but rather is a  special feature of  semiclassical 
expansion near a certain   string configuration (represented by 
massless \adss geodesic  wrapping $S^5$). 
Expanding near other string solutions leads to other 
special ``Penrose-type'' limits of string geometry (geometry ``seen'' by a
fundamental string probe)  which are described by the
corresponding quadratic fluctuation actions. 
Expansions near  a class of ``fast'' string configurations  (discussed below) 
 for which 
the world sheet  becomes null in the effective  zero-tension limit
 \ci{mateos,mik,mikk} may be interpreted as a stringy generalization of 
the  Penrose limit (keeping in mind, of course, that this analogy applies 
only in the strict zero-tension limit).} 
 and then of \ci{gkp} 
that a progress  can be made by 
 (i) concentrating on a subsector of states with large 
or ``semiclassical''  values 
of quantum numbers, $Q \sim T \sim \sql$
%(here $Q$ stands for a generic quantum number like spin 
%in $AdS_5$ or $S^5$ or an oscillation number)
and (ii) considering a  new limit 
\be  \la{li}
Q \to \infty \ , \ \ \ \ \ \ \ \ \  \ \ \ \ \ \ 
\tl \equiv {\l \ov Q^2} = {\rm fixed}   \ .
\ee 
On the string side ${Q \ov \sql} = { 1 \ov \sqrt{ \tl}}$ 
plays the role of a  semiclassical parameter 
(like rotation frequency) which can then be taken to be large. 
The energy of such states  is 
$E= Q + f(Q,\l)$, where  $f\to 0$ in the $\l \to 0$ limit. 
The duality implies that such semiclassical string states   
(as well as states represented by small  fluctuations near them) 
 should be dual  to 
``long'' SYM operators with a large canonical dimension, 
i.e.  containing large number of fields or derivatives 
under the trace.  In this case the duality map becomes more explicit.

The simplest possibility  is to start with a BPS state that 
carries a large quantum number  and consider  small fluctuations 
 near it, 
 i.e. a set of {\it near-BPS}  states   characterised 
by   a large parameter \ci{bmn}.
The only non-trivial example of such a BPS state is 
represented  by  a 
point-like string moving along  a geodesic in $S^5$ with a 
large angular
 momentum $Q=J$.
Then $E=J$ and the dual operator is $\tr\Phi^J$,\ \  $\Phi= \p_1 + i \p_2$. 
The small 
closed strings representing near-by  fluctuations 
  are ultrarelativistic, i.e.  their  kinetic energy 
is much larger  than  their mass.
They 
are dual to SYM operators of the form  $ \tr( \P^J ...)$ 
where 
dots stand for a small number of other fields 
and/or covariant derivatives
(one needs to sum over different orders of the factors
 to find an eigenstate of the anomalous dimension matrix). The 
energies of the  small fluctuations  happen to have the structure  
 \ci{mets,bmn}
 \be \la{ytr}
 E= J +  \sqrt{ 1 + n^2 \tl }\  K +  O({1\ov J})\ =
 J +  K +  k_1 \tl + k_2 \tl^2 + ...   \ . \ee 
One can argue in general \ci{ft1,tse1} and check explicitly \ci{par,cal}
that higher-order quantum string sigma model corrections
are indeed suppressed in the limit \rf{li}, i.e. in the 
 large $J$, fixed $\tl\equiv  { \l \ov J^2} = \l'$ limit. 
A remarkable feature of this expression is 
that  $E$ is analytic in $\tl$, suggesting direct comparison with 
perturbative  SYM expansion in $\l$. 

Indeed, it can be shown 
that the first three  $\tl$, $\tl^2$  and $\tl^3$ 
 terms in the expansion 
of the square root agree precisely 
with the one \ci{bmn},  two \ci{grom} 
and three \ci{bks,beit,soc} 
loop terms in the anomalous dimensions of the corresponding operators.
There is also (for a 2-impurity $K=2$ case)  
 an   argument \ci{zan} 
suggesting  how the full  $\sqrt{ 1 + n^2 \tl }$ expression 
may  appear
on the perturbative SYM side (for reviews of various  aspects 
 of the BMN limit see also \ci{beb}).
However, the general proof of the consistency 
of the BMN limit in  the SYM theory  (i.e. 
that the usual perturbative expansion 
can be rewritten as an expansion in $\tl $ and $1 \ov J$) 
remains to be given.
Also, 
 to explain why the string and the SYM expressions match 
 one should 
show that the string limit  (first $J\to \infty$, then
  $\tl= { \l \ov J^2}\to 0$)  and  the SYM limit 
(first $\l \to 0$, then $J \to \infty$)  produce exactly 
the same expressions for the energies/dimensions, 
even though, in general, 
the two limits may not commute, cf. \ci{kvs,SS,bds}. 

If one moves away from the near-BPS limit  and considers, e.g.,  
a non-supersymmetric  state with a large angular momentum 
$Q=S$ in $AdS_5$
\ci{gkp},  a similar   direct 
 quantitative check of the duality is no longer possible:
here the classical energy is not analytic in $\l$ and 
quantum corrections are no longer suppressed by powers of $1\ov S$
(but as usual  are  suppressed by powers of $1 \ov \sqrt\l$).
However, it is still possible to demonstrate a remarkable qualitative 
agreement between $S$-dependence 
of the string energy and of the SYM anomalous dimension.
The energy  of a folded closed string rotating  at the center of 
$AdS_5$  which is dual \ci{gkp} 
to  twist 2 operators on the SYM side
($\tr ( \Phi^*_k D^S \Phi_k), \ \ D= D_1 + i D_2,$ and similar operators 
with spinors and gauge bosons that mix at higher loops \ci{klv,klvo})
has the following form when  expanded at large $S$ : \be 
E= S + f(\l)  \ln S +  O(S^0) \  \ee
  On the string side \be \la{lar} 
  f(\l)_{_{\l \gg 1 }} = c_0 \sql + c_1 + {c_2 \ov \sql} + ...\ ,\ee
where $c_0= { 1 \ov \pi}$ is the classical \ci{gkp} and 
$c_1=  - { 3 \ov  \pi} \ln 2$  is the 1-loop \ci{ft1}  coefficient. 
On the gauge theory side one finds the {\it same} $S$-dependence 
of the anomalous dimension with the 
perturbative expansion of the $\ln S$ coefficient being \be \la{sma}
f(\l)_{_{\l \ll 1 }} = a_1 \l + a_2 \l^2  + a_3 \l^3  + ...\ \ee
where 
$a_1= { 1 \ov 2 \pi^2} $ \ci{gw}, $a_2= - { 1 \ov 96 \pi^2} $ \ci{klv}, 
and $a_3=  { 11 \ov 360 \times 64  \pi^2} $ \ci{klvo}.
Like in the case of the SYM entropy \ci{gkt}, here one expects the existence
 of a smooth interpolating function $f(\l)$ that connects the two 
perturbative expansions. 
In fact, observing that the factor $1\ov \pi$ in \rf{lar} 
and factor $1 \ov \pi^2$ in \rf{sma}  seem to factorize, 
one can suggest a simple square root type interpolating formula 
for $f(\l)$ that seem indeed to give a good fit \ci{klv,klvo}
(cf. also the discussion  end of section 4).

%%%%%%%%%%%%%%%%%%%%%%%%%%%%%%%%%%%%%%%%%%%%%%%%%%%%%%%%%%%%%
\subsection{Multispin string states} 
%%%%%%%%%%%%%%%%%%%%%%%%%%%%%%%%%%%%%%%%%%%%%%%%%%%%%%%%%%%

One may    wonder still if  examples of quantitative agreement 
between string energies and SYM dimensions  found  for the near-BPS 
 (BMN) states exist  also for more general non-BPS string states. 
Indeed, it was noticed  already  in \ci{ft1} that a string state 
that carries large spin in $AdS_5$ as well as large spin $J$ 
in $S^5$ has, in contrast to the above $J=0$ case, 
 an analytic expansion of its energy  in $\tl= {\l \ov J^2}$, 
just as in the BMN case with a large  oscillation number  $K \sim S$. 
 It was observed  in \ci{ft2} that  semiclassical 
 string states carrying several large spins (with at least one of them
being in $S^5$)  have a 
regular expansion of their energy $E$
in  powers of $\tl$   and it was then  suggested, 
 by analogy with the  near-BPS  case, that the expansion of $E$ 
 in small effective tension  or $\tl$ 
may be possible to match with the 
perturbative expansion of  the SYM dimensions. 

For a classical  rotating closed string solution in $S^5$  one has 
$E= \sql {\mathcal E}(w_i) , \  J_i = \sql w_i$ so that 
$E=E(J_i, \l)$. The  required key  property is that (in contrast to  the case of a single spin 
in $AdS_5$) there should be  no 
$\sql$ factors in the expansion  of the classical energy 
$E$  in small $\l$ 
\be \la{hoi} 
E= J + c_1 { \l \ov J} + c_2 { \l^2 \ov J^3} + ...
= J \ \left[ 1 + c_1 \tl + c_2 \tl^2 + ...\right] \ .\ \ee
Here $J= \sum_{i=1}^3 J_i, \ \tl \equiv {\l \ov J^2}$
and $c_n= c_n({J_i \ov J})$ are functions of
  ratios of the spins which are finite in the limit 
$ J_i \gg 1$, $\tl =$fixed. 

 The  simplest  example of such  solution is 
provided by a circular string rotating in two orthogonal planes in $S^3$ part of $S^5$
with the two angular momenta being equal,  $J_1=J_2$  \ci{ft2}:
$$\XX_1 \equiv X_1 + i X_2 = \cos (n \s) \ e^{i w \tau}\ , \ \ \ \ 
\XX_2 \equiv X_3 + i X_4 = \sin (n \s) \ e^{i w \tau} \ , $$
and 
with the global $AdS_5$ time being 
$t= \k \tau$.
The conformal 
gauge constraint implies 
$\k^2= w^2 + n^2$ and thus 
\be \la{eee}
E= \sqrt{ J^2  + n^2 \l} = 
J ( 1  + \ha  n^2 \tl - { 1 \ov 8 } n^4 \tl^2 + ...) \ ,  \ee
where $J= J_1+J_2=2 J_1$.
 For  fixed $J$ the  energy thus has a  regular 
expansion in string tension (in contrast to what happens 
in flat space  where  $E= \sqrt{ { 2 \ov \a'} J}$). 

Similar expressions \rf{hoi} are found also for
 more general rigid multispin 
closed  strings \ci{ft2,ft3,ft4,afrt,art,tse2}. 
In particular,  for a folded  string rotating in one plane of
 $S^5$ and with its center of mass orbiting along big circle in another
 plane 
the coefficients $c_n$ are transcendental functions 
(expressed in terms of elliptic integrals) \ci{ft4}. 
More generally, the 3-spin solutions are described by an 
integrable Neumann model \ci{afrt,art}  and  the coefficients 
$c_n$ in the energy are expressed in terms of genus two 
hyperelliptic functions. 
The reason why  choosing  a particular 
string ansatz one gets  an  integrable effective 1-d model
 lies  in the 
integrability  of the original $S^5= SO(6)/SO(5)$  classical  
sigma model \ci{pohl} (see also \ci{as}).

To be able to hope to compare the classical energy  to the SYM 
dimension one should be sure that 
higher  string $\a'$  corrections 
are suppressed in the limit $J \to \infty, \ \tl=$fixed. 
Formally, this is of course   the case since $\a' \sim { 1 \ov \sql} 
\sim { 1 \ov J \sqrt{ \tl}}$; 
what is more important, the $1 \ov J $ corrections are again 
analytic in $\tl$
\ci{ft3},  i.e., as in the BMN case, 
 the expansion in large $J$ and small $\tl$ 
is well-defined on the string side, 
\be \la{ere}
E= J \ \left[ 1 + \tl ( c_1 + { d_1 \ov J} + ...) 
+  \tl^2 ( c_2 + { d_2 \ov J} + ...)  + ... \right] \ , \ee
with the classical energy \rf{hoi} being the $J\to \infty$ limit 
of the exact expression. 

The reason for   this  particular form of the energy  \rf{ere}  can be explained 
as follows \ci{ft1,tse1,tse2}. We are computing the \adss 
 superstring sigma model 
loop corrections to the mass of a stationary solitonic solution on a 
2-d cylinder (no IR divergences). This theory is conformal 
(due to the crucial presence of the fermionic fluctuations) and thus
it does not depend on a  UV cutoff. 
The relevant 2d fluctuations are massive and their
 masses scale as $w \sim 
{1 \ov \sqrt{\tl}}$. As a result,  the inverse mass expansion 
is well-defined and the quantum corrections should be proportional 
to positive powers of $\tl$.
This was explicitly demonstrated  by a 1-loop computation in 
\ci{ft3,fpt}. 

Similar expressions are found  for the energies of small
fluctuations near a given classical solution: as in the BMN case, 
the fluctuation energies are suppressed by extra  factor of $J$, 
i.e. 
\be \delta E = \tl ( k_1 + { m_1 \ov J} + ...) 
+  \tl^2 ( k_2 + { m_2 \ov J} + ...)  + ... \ . \ee

%%%%%%%%%%%%%%%%%%%%%%%%%%%%%%%%%%%%%%%%%%%%%%%%%%%
\subsection{AdS/CFT duality:  non-BPS states}
%%%%%%%%%%%%%%%%%%%%%%%%%%%%%%%%

Assuming that the same large $J$ 
 limit is well-defined also on the SYM side, 
one should then be  able to compare the coefficients in \rf{ere}
to the coefficients  in the anomalous 
dimensions of the corresponding SYM operators  \ 
$\tr( \P_1^{J_1} \P_2^{J_2}  \P_3^{J_3}  ) + ...$
(and also do similar matching for near-by fluctuation modes) \ci{ft2}. 
In practice, what  is known (at least in principle) is  
how to compute the dimensions in the  different limit: 
by first expanding in $\l$  and then expanding in $1 \ov J$.
One may expect that this expansion  of anomalous dimensions 
may take the form equivalent to \rf{ere}, i.e.
\be \la{dela}
\D= J + \l ( { a_1 \ov J} + { b_1 \ov J^2} + ...) 
      + \l^2  ( { a_2 \ov J^3} + { b_2 \ov J^4} + ...) + ...\ , \ee 
and that  the respective coefficients in $E$
and $\D$  may agree with each other. 
The subsequent work \ci{mz1,bmsz,bfst,as,Min2,kru,SS,kmmz,krt,zarL}
did  verify this  structure of $\D$ and,  moreover,  
established the general agreement 
between the two leading coefficients $c_1,c_2$  in $E$ \rf{ere}
and the one-loop and two-loop coefficients 
$a_1,a_2$ in $\D$ \rf{dela} 
(as usual, by ``$n$-loop'' term in $\D$  we mean
 the term 
multiplied by $\l^n$).

To compute $\D$  one is   to  diagonalize the 
anomalous dimension matrix defined on a set of ``long'' scalar 
operators, and this   is obviously a non-trivial problem.
The  important  step to this goal 
 was made in \ci{mz1} where it was 
observed that the one-loop planar  dilatation operator in the scalar sector 
can be interpreted as a Hamiltonian of an integrable $SO(6)$
 spin chain and thus can be diagonalized even for large 
length $L=J$  by the  Bethe ansatz method.\foot{Relations 
between 1-loop anomalous dimension  matrix for a  certain class of 
 composite operators  and   
integrable spin chain Hamiltonians  were observed previously in the 
large $N$ QCD context \ci{qcd} (for a review and connections 
to AdS/CFT   see \ci{kog}).}
In the simplest case of the  ``$SU(2)$'' sector  of operators 
$\tr( \P_1^{J_1} \P_2^{J_2}) + ...$  built out of 
two chiral scalars,  the dilatation operator 
  can be interpreted as ``spin up'' and 
``spin down'' states of periodic XXX$_{1/2}$  spin chain with length 
$L=J=J_1+J_2$. Then 
the 1-loop dilatation operator becomes equivalent 
to the Hamiltonian of the ferromagnetic Heisenberg model 
\be \la{ferr}
D_1 = { \l \ov (4 \pi)^2} \sum^J_{l=1}(I - {\vs}_l \cdot { \vs}_{l+1}) 
 \ . \ee
By considering the thermodynamic 
limit ($J \to \infty$) of the  corresponding Bethe ansatz equations, 
the 
proposal of  \cite{ft2} was confirmed at the leading order 
of expansion in $\tl$  \cite{bmsz,bfst}: 
for eigen-operators  with $J_1\sim J_2 \gg 1 $ 
it was shown  (i)  that 
$\D-J=  \l {a_1 \ov J} + ...$,  and  (ii) a  
 remarkable agreement 
was found between $a_1=a_1({J_1 \ov J_2})$ and the coefficient $c_1$
 in the energies \rf{ere}  
 of  various 2-spin  string solutions.
It was also possible also to match 
(as in the BMN case) 
the energies of  fluctuations 
near the circular $J_1=J_2$ solution 
with the corresponding eigenvalues 
of \rf{ferr} \ci{bmsz}.

Similar leading-order agreement between string energies and 
SYM dimensions was observed also in other sectors of states
with large quantum numbers: 

(1) in  the $SU(3)$ sector:  for  specific 
solutions \ci{ft2,afrt,art} 
 with 3 spins in $S^5$  which are dual to the operators  $\tr( \P_1^{J_1}
 \P_2^{J_2}  \P_3^{J_3}  ) + ...$ 
\ci{Min2,char}; 

(2)  in the $SL(2)$ \ci{BS} sector:
for a folded string state \ci{ft1} 
with one spin in $AdS_5$ and one spin in $S^5$
 (with $E=
J + S + { \l \ov J} c_1 ( {S\ov J}) + ...$ \ci{ft1,ft2}) 
which is dual to the operators 
$\tr ( D^S \P^J) + ... $ \ci{bfst};

(3) in a ``subsector'' of $SO(6)$ states 
containing pulsating (and rotating)  solutions 
which again have regular \ci{Min1} expansion of the energy 
  in the limit 
of large oscillation  number $L$, i.e. 
 $E=L + c_1 { \l \ov L}  + ...$  \ci{bmsz,Min2}.
 
This agreement between the leading-order terms in the 
expansion of energies of  certain 
semiclassical string states and  dimensions of the corresponding 
``long'' SYM operators leaves, however, 
 many questions, in particular:

(i) How to understand this agreement  beyond specific examples, 
i.e. in a more universal way? 

(ii) Which is the  precise relation between profiles of string 
solutions and the structure of the dual SYM operators?

(iii) How to characterise the set of semiclassical string states 
and dual SYM operators to  which this direct 
relation  should apply?

(iv) Why the agreement holds at all, i.e. why the two limits 
(first $J\to \infty$, and then $\tl \to 0$, or vice versa) 
taken on the string and the SYM sides 
give equivalent results to the first two orders in expansion in $\tl$? 
Why/when it does not 
 work to all orders in expansion in $\tl$ (and $1\ov J$)? 

 The questions (i),(ii) were addressed 
in \ci{kru,krt,lopez,ST,kt} using the low-energy effective action approach
for coherent states;
an alternative approach based on matching 
the general solution (and the integrable structure) 
of the  string sigma model with that of the thermodynamic limit 
of the Bethe ansatz in the $SU(2)$ sector was developed in \ci{kmmz}.
The question (iii) was addressed  in 
\ci{mateos,mik,mikk,kt}, and  the question (iv) -- in 
\ci{SS,bds,afs}. Still,  our understanding  of why 
there is a  direct agreement  with gauge theory at the first 
 two  $\tl$ and $\tl^2$ orders of  expansion  and why it does not
  \ci{bfst,bds}
 continue to the $\tl^3$ order is still
  rather rudimentory.\foot{Similar (dis)agreements  were found  for the 
  $1/J$ corrections to the BMN states \ci{cal}.}

 Below we shall review the effective action approach
 as developed in \ci{kru,krt,ST,kt}, concentrating mostly 
 on  the leading (``1-loop'') order in expansion in $\tl$.

%%%%%%%%%%%%%%%%%%%%%%%%%%%%%%%%%%%%%%%%%%%%
\setcounter{equation}{0}
%%%%%%%%%%%%%%%%%%%%%%%%%%%%%%%%%%%%%%%%%%%%
%\setcounter{footnote}{0}
\renewcommand{\theequation}{\thesection\arabic{equation}}
%\renewcommand{\thefootnote}{\fnsymbol{footnote}}

%%%%%%%%%%%%%%%%%%%%%%%%%%%%%%%%%%%%%%%%%%%%%%%%%%%%%%%%%%
\section{Effective actions for\\  coherent states}
%%%%%%%%%%%%%%%%%%%%%%%%%%%%%%%%%%%%%%%%%%%%%%%%%%%%%%
%%%%%%%%%

The  suggestion  of how to  understand  the agreement between 
leading-order terms in the multispin 
string energies and the corresponding one-loop  anomalous 
dimensions in a universal way 
was made  in  \ci{kru}  and was clarified and 
elaborated further  in \ci{krt,kt}.
 The key idea   was that instead of comparing particular 
solutions one should try to match  effective sigma models 
which appear on the string side  and the SYM side.
Another related   idea of \ci{kru,krt,kt} 
was that since the ``semiclassical'' string states 
carrying  large quantum numbers are represented in the 
quantum theory  by coherent states, one should be comparing coherent 
string states to coherent SYM states (and thus  to coherent 
states of the spin chain).
In view  of the ferromagnetic 
nature of the dilatation operator \rf{ferr},  in   the thermodynamic 
limit 
$J=J_1+J_2  \to \infty$ with fixed  large  number of impurities
(i.e. with  fixed  $J_1 \ov J_2$) 
it is  favorable to form large clusters of spins. Then  a 
``low-energy''  
approximation and continuum limit should 
 apply, leading to an effective 
``non-relativistic''  sigma model for a coherent-state expectation 
value of the spin operator. 

At the same time, on  the string side, 
taking  the ``large space-time energy'' (or large $J$) 
  limit directly 
in the classical string action produces   a reduced ``non-relativistic'' 
sigma model that describes in a universal way 
the  leading-order $O(\tl)$ corrections 
to  energies of all string solutions in the two-spin sector.
The resulting sigma model action turns out to 
agree  exactly \ci{kru}  with   the semiclassical 
 coherent state action found from the 
$SU(2)$ sector of the  spin chain 
in the $J \to \infty, \ \tl=$fixed  limit. 
This demonstrates 
how a string action can directly ``emerge'' from a gauge 
theory in the large-$N$ limit and 
provides a direct map between  the 
``coherent''  SYM  states (or the corresponding 
operators built out of  two holomorphic  
scalars) and all two-spin classical string states.
Furthermore, the correspondence  established at the level
of the action implies also (i) the matching of the integrable structures 
and (ii)  the matching of the fluctuations 
around particular  solutions
 and thus  it 
   goes  beyond the special examples of 
 rigidly rotating strings. 

%%%%%%%%%%%%%%%%%%%%%%%%%%%%%%%%%%%%%%%%%%%%%%%%%
\subsection{Coherent states}
%%%%%%%%%%%%%%%%%%%%%%%%%%%%%%%%%%%%%%%%%%%%%%%

Let us briefly 
review the definition of coherent states
(see, e.g.,  \ci{pere}). For a harmonic oscillator  
($ [a,a^\dagger]=1$) one can define  the coherent state 
 $\ket{u}$ as
$a \ket{u} = u \ket{u}$, where $u$  is a complex number. 
Equivalently, $\ket{u}= R(u) \ket{0}$, where 
$R= e^{ u a^\dagger - u^* a} $ so that acting on the vacuum 
$\ket{0}$ the operator 
$R$ is simply proportional to $e^{ u a^\dagger} $.
Note that $\ket{u}$  can be written as a  superposition 
of the  eigenstates $\ket{n}$ of the harmonic oscillator Hamiltonian, 
$\ket{u} \sim \sum^\infty_{n=0} { u^n \ov \sqrt{ n !}}
\ket{n}$.
An alternative definition of a  coherent state is  that it is a state 
with minimal uncertainty for both the coordinate 
$\hat q = { 1 \ov \sqrt 2} ( a +  a^\dagger) $ and the 
momentum $\hat p = -{ i \ov \sqrt 2} ( a - a^\dagger) $
operators, $\Delta \hat p^2 = \Delta \hat q^2 = \ha$, 
\ $ \Delta \hat p^2 \equiv \bra{u } \hat p^2  \ket{u}
- ( \bra{u } \hat p \ket{u})^2 $. 
For that reason it is the ``best''
 approximation to a classical state.
If one defines a time-dependent state $\ket{u(t)}
= e^{- i H t} \ket{u}$ then the expectation values of 
$\hat q$ and $\hat p$, i.e.  
$\bra{u } \hat q \ket{u} = { 1 \ov \sqrt 2} ( u + u^*), \ 
$ $ \bra{u } \hat p \ket{u} =- { i \ov \sqrt 2} ( u -u^* )$
 follow the classical trajectories.

 Starting instead of the Heisenberg algebra 
 with the $SU(2)$ algebra 
$[S_3, S_\pm ] = \pm S_\pm, \ [S_+, S_-] = 2 S_3$ 
and considering the $s=1/2$ representation where
 $\vec S = \ha \vec \s$ 
one can define a  spin coherent state
 as a 
linear superposition of spin up and spin down
states: $\ket{u} = R(u) \ket{0}. $ Here 
$ R= e^{ u S_+ - u^* S_-}$,  $ \ket{0} = \ket{\ha, \ha}$
and $u$ is a complex number.
% that can be parametrized as 
%$u= \ha \theta e^{i \p}$. 
An equivalent way to 
label  the  coherent state is  by a unit 3-vector $ \vec n$ 
defining a point of $S^2$. Then $ \ket{\vec n }
= R( \vec n) \ket{0 }$ where $ \ket{0 } $ 
corresponds to a 3-vector $(0,0,1)$ along the 3rd axis.
One can write  $ \vec n= U^\dagger \vec \s U, \ U= (u_1,u_2)$,   
and then  $R( \vec n)$ is an $SO(3)$ rotation  from a north pole to 
a generic point of $S^2$ defined by $\vec n$.
  The key property of the coherent state is 
that $\vec n$ determines the coherent state 
expectation value of
 the spin operator:
\be \la{coha}
\bra{\vec n  } \vec S  \ket{\vec n} = \ha  \vec n\  . \ee
Similar definition of coherent states 
can be given in the case when $SU(2)$ is replaced by 
a  generic  group $G$. 
Given a semisimple group $G$  with  the Cartan basis 
of its algebra  $(\rH_i,\rE_\a,\rE_{-\a})$ 
($[\rH_i,\rH_j]=0, \ [\rH_i,\rE_\a ]=\a_i \rE_\a  , \ 
[\rE_\a, \rE_{-\a} ]= \a^i \rH_i , \ 
[\rE_\a, \rE_{\b} ]= N_{\a\b} \rE_{\a+\b}$) 
whose interpretation will be 
 a symmetry group of a quantum  Hamiltonian (acting 
in a  unitary irreducible representation $\L$ on the Hilbert 
space $V_\L$)
one may define a set of coherent states
 by choosing a particular state 
$\ket{0}$ (with $ \bra{0}\ket{0}=1$)  in $V_\L$
and acting on it  by the elements of $G$. A subroup $H$ of $G$ that
leaves 
$\ket{0}$ invariant up to a phase 
($ \L(h)\ket{0} = e^{i\phi(h)} \ket{0}$) 
 is called maximum stability subgroup.
One may then define the coset space $G/H$, the  elements  of  which 
($g = \omega h, \ h \in H, \ \omega \in G/H$, \ $\L(g)
=\L(\omega)\L(h)$) 
will parametrize the coherent states, 
$\ket{\omega,\L} = \L(\omega) \ket{0}$.

This definition depends on a choice of group $G$, its
representation 
$\L$ and the vector   $\ket{0}$. 
It is natural to assume also 
that $\ket{0}$ is an eigenstate 
 of the Hamiltonian $H$, e.g., a ground state. 
 For a unitary representation $\L$ we may choose 
 $\rH_i^\dagger=\rH_i  ,\ \rE_\a^\dagger = \rE_{-\a}$
 and select   $\ket{0}$ to be the highest-weight vector
  of the 
 representation  $\L$, i.e. demand that it is annihilated by
 ``raising'' generators  and is an eigen-state of the Cartan 
 generators: 
 (i) $ \rE_\a \ket{0}=0$ for all positive roots $\a$; \ \ 
 (ii) $\rH_i  \ket{0} = h_i  \ket{0}$.
 In addition, we may demand that $ \ket{0}$ is annihilated also 
 by some ``lowering'' generators, i.e. 
 (iii)  $ \rE_{-\b} \ket{0}=0$ for {\it  some}  negative 
  roots $\b$; the remaining negative roots will be denoted by $\g$. 
  Then  the  coherent states are given by   
  \be 
\ket{\omega,\L} = {\rm exp}\big[ \sum_\g (w_\g  E_{-\g} -
  w^*_\g  E_{\g} )\big]\ 
  \ket{0}\ , 
\ee
  where $\g$ are the negative roots for which $\rE_\g \ket{0}
  \not=0$.  $w_\g$ may be interpreted as coordinates 
  on $G/H$ where $H$ is generated by $(\rH_i,\rE_\a,\rE_{-\b})$.
  
  For example, in the case of $G=SU(3)$ 
  with the  Cartan basis
  
  \noindent 
  $ (\rH_1,\rH_2, \rE_\a,\rE_{\b}, \rE_{\a+\b},
  \rE_{-\a},\rE_{-\b}, \rE_{-\a-\b})  $
  and 
  with $\ket{0}$ 
  being the highest-weight of the fundamental representation, 
  i.e. $\rE_{-\b}\ket{0}=0$, $\rE_{-\a}\ket{0}\not=0$, 
  $\rE_{-\a-\b}\ket{0}\not=0$,
  the subgroup $H$ is generated 
  by $(\rH_1,\rH_2, \rE_\b,\rE_{-\b})$, i.e. is $SU(2)\times U(1)$
  and $G/H = SU(3)/(SU(2)\times U(1))= CP^2$
  (see also \ci{ST}). 
  We shall apply this general definition of coherent states 
  in section 3.

%%%%%%%%%%%%%%%%%%%%%%%%%%%%%%%%%%%%%%%%%%%%%%%%%
\subsection{Landau-Lifshitz model \\ from spin chain}
%%%%%%%%%%%%%%%%%%%%%%%%%%%%%%%%%%%%%%%%%%%%%%%

In general, one can rewrite the usual phase space path integral 
as an integral over the overcomplete set of coherent states 
(for the harmonic oscillator this is 
simply a  change of variables from $q,p$ to 
 $ u = { 1 \ov \sqrt 2} ( q + i p)$):
\be\la{zz}
 Z= \int [du] \ e^{i \S[u]}  \ , \ \ \ \ \ \ \ \ \ 
\S= \int dt \bigg( \bra{u }  i  { d \ov dt} \ket{u}  - 
\bra{u }  H  \ket{u} \bigg) \ . \ee
 The first (``Wess-Zumino'' or ``Berry phase'') 
term in the action $\sim i u^* { d \ov dt} {u}$
is the analog of the usual $ p \dot q$ term
 in the phase-space action. Applying this to the case of the 
Heisenberg 
spin chain Hamiltonian 
\rf{ferr} one ends up with  with the following action 
for the coherent state variables $\vec n_l(t)$ at sites $l=1,...,J$
 (see, e.g.,  \ci{fra}):
\be \S=\int dt \sum^J_{l=1}  \bigg[ \vec C (n_l) \cdot \vec n_l  
- { \l \ov 2 (4 \pi)^2 }  ( \vec n_{l+1} 
- \vec n_l)^2  \bigg] \ . \ee
Here $d C= \epsilon^{ijk} n_i d n_j \wedge d n_k$
(i.e. $\vec C$ is a monopole potential on $S^2$). 
In local coordinates (at  each site $l$) one has 
$\vec n = (\sin \theta \ \cos \p, \ 
 \sin \theta \ \sin \p,\ \cos \theta)$, 
\ $\vec  C \cdot d \vec n = \ha \cos \theta  d \p$.
In the limit $J\to \infty$  with fixed $\tl= {\l \ov J^2}$ 
(which we are interested in) 
 we  can take  a continuum limit 
by introducing the 2-d field $ \vec n(t,\s)= \{ \vec n (t,
 { 2 \pi \ov J} l) \}$, $l=1,...,J$.
  Then 
\be \la{con}
\S= J \int dt \inti  \left[ \vec C \cdot  \del_t  \vec n - 
{1 \ov 8} \tl (\del_\s \vec n)^2  + ... \right] \ , \ee
where dots stand for higher derivative terms suppressed by 
$1 \ov J$. Since $J$ appears as a factor in front of the action, 
in  the limit $J \to \infty$  
 all quantum corrections should be also   suppressed 
by $1 \ov J$, and thus the above action can be treated classically.
The corresponding equations of motion 
\be  \del_t  n_i = \ha \tl \epsilon_{ijk} n_j \del^2_\s n_k \ee 
 are the 
Landau-Lifshitz equations
for a classical ferromagnet. 
An alternative  derivation  of them 
is based on first writing down  the 
Heisenberg equation for the time evolution of the 
spin operator 
directly from the spin chain Hamiltonian,  
 then considering  the coherent  state expectation value 
and finally taking the continuum  limit.

%%%%%%%%%%%%%%%%%%%%%%%%%%%%%%%%%%%%%%%%%%%%%%%%%
\subsection{Landau-Lifshitz model \\ from string action}
%%%%%%%%%%%%%%%%%%%%%%%%%%%%%%%%%%%%%%%%%%%%%%%

The action \rf{con} should be describing the coherent states 
of the Heisenberg spin chain in the above thermodynamic limit. 
One may wonder how a similar ``non-relativistic'' 
action may appear  on the string side where one starts 
with the  usual ``relativistic'' sigma model \rf{ss}. 
 To obtain such an  effective action 
 one  is  to 
perform the following sequence of steps  \ci{kru,krt,kt}:

(i) isolate a  ``fast'' coordinate $\a$ whose momentum  $p_\a$ is
large for a class of string configurations we consider; 

(ii) gauge-fix $t \sim \tau$ and 
$ p_\a \sim J$ (or $\td \a \sim \s$ where $\td \a $ is ``T-dual''
to $\a$); 

(iii) expand the action in  derivatives of 
``slow'' or ``transverse'' coordinates 
(to be identified with $\vec n$).

To illustrate this procedure let 
 us consider  the $SU(2)$ sector of string states 
carrying two large spins in $S^5$, with string motions restricted 
to $S^3$ part of $S^5$. The relevant part of the \adss metric is then 
$ds^2= - dt^2 + d\XX_i d\XX_i^*$, with $\XX_i \XX_i^*=1$.
Let us 
set
 $$\XX_1 = X_1 + i X_2 = u_1 e^{i \a}\ , \ \ \ \ \ \ \ \ \ 
\XX_2 = X_3 + i X_4 = u_2 e^{i \a}\ , \ \ \ \ \ \ \ \ 
 u_i u^*_i=1\ . $$
Here $\a$ will be a  coordinate  associated to the 
total  spin in the two planes (which in general 
will be the sum of orbital  and internal spin).  
$u_i$   (defined modulo $U(1)$ 
gauge transformations) will be the  ``slow'' 
coordinates determining the
 ``transverse'' string profile. Then 
$$d\XX_i d\XX_i^* = ( d \a + C)^2 + D u_i Du^*_i   , \ \ \ \ \ \  
C= - i u^*_i du_i , \ \ \ \ \ \ \ Du_i = d u_i -i C u_i  ,  $$
 where   the second  $ |Du_i|^2$ term 
represent the metric of $CP^1$ (this parametrisation corresponds 
to the  Hopf fibration $S^3 \sim S^1 \times S^2$). 
Introducing $\vec n = U^\dagger \vec \s U, \ U=(u_1,u_2)$ 
we get 
\be 
d\XX_i d\XX_i^* = (D \a)^2 +  { 1 \ov 4} ( d \vec n)^2 \ , 
\ \ \ \ \ \ \ \ \ \ \  
\ D\a= d \a + C(n) \ .\ee 
 Writing the resulting sigma model action in phase space form 
and imposing the (non-conformal) 
gauge $t= \tau, \ p_\a =$const$=J$,  
one gets \ci{krt} 
the same  action \rf{con} with the WZ term $
\vec C \cdot  \del_t  \vec n$
originating from the  $p_\a D \a $ term in the phase-space 
Lagrangian (cf. its origin on the spin chain  side
as an analog of the  $p \dot q$ in the coherent state 
path integral action). 

An equivalent approach \ci{kt} 
 leading to the same action \rf{con}
is based on first applying a 2-d 
duality (or ``T-duality'') transformation 
 $\a\to \td \a $  and then choosing the 
``static'' gauge $t= \tau, \ \  \td \a ={ 1 \ov \sqrt{ \td \l}  } \s$
with $    { 1 \ov \sqrt{ \td \l}  }= { J \ov \sql}$.  
Indeed, 
starting with \be
\LL =  - \ha \sqrt{ - g}\ g^{pq}
\big( - \d_p t \d_q t + D_p \a  D_q \a  + D_p u^*_i 
D_q u_i
\big) \ee and 
applying  the 2-d duality in $\a$ we get 
\be\LL =  - \ha \sqrt{ - g} g^{pq}
\big( - \d_p t \d_q t + \del_p \ta  \del_q \ta  + D_p u^*_i
 D_q u_i
\big)  +   \ep^{pq} C_p \del_q \ta \ . \ee
Thus the ``T-dual''  background  has no off-diagonal
 metric component 
but has a  non-trivial NS-NS 2-form  coupling 
in the $(\ta,u_i)$  sector.  
It is useful   not to use
conformal gauge here. 
Eliminating the  2-d metric $g^{pq}$ we then get 
the Nambu-type action   
\be \la{namb} 
\LL  =     \ep^{pq} C_p \del_q \ta\  - \sqrt{  h } 
\   ,  \ee
where 
$
h =|\det \  h_{pq}| $ and 
$ h_{pq} = 
- \d_p t \d_q t + \del_p \ta  \del_q \ta 
 + D_{(p}  u^*_i D_{q)} u_i$. 
If we now   fix the  static gauge,  
$
t = \tau , \  \ta=  {1 \ov  \sqrt{ \td \l}  } \s $, 
we finish with the  action 
$I =  J \int dt \inti \ \LL$, where 
\be 
\la{eaa}
\LL  =  C_t  -  \sqrt{  (1+  \tl  |D_\s u_i|^2)( 
1 -  |D_t u_i|^2  )
+  \fo \tl   (D_t
 u^*_i D_\s u_i + c.c.)   ^2 }\ . 
\ee
Making the key assumption that the evolution of $u_i$ 
in $t$ is slow, i.e. the time derivatives are suppressed
(which can be implemented by  rescaling $t$ by $\tl$ and 
expanding in powers of $\tl$), 
we find, to the  leading order in $\tl$,  
\be \la{coon}
\LL = - i u^*_i \del_t u_i - \ha \tl |D_\s u_i|^2 \ . \ee
        This  is the same as the $CP^1$ Landau-Lifshitz 
action  
 \rf{con} when written in terms of 
$\vec n$. 
Thus the string-theory  
counterpart   of the WZ term   in the spin-chain
coherent state  effective 
 action  originates  from the 2-d  NS-NS WZ term in the action for the 
``T-dual''  coordinate  $\ta$ upon
the  static gauge fixing of the latter   \ci{kt}. 

To summarize: 

(i) $(t, \ta)$ are the  ``longitudinal'' coordinates 
that are gauge-fixed (with $\ta$ playing 
the role of the  string direction or the  spin chain direction on the
SYM side);

 (ii)  $U=(u_1,u_2)$  or $\vec n = U^\dagger
\vec \s U$ are  the ``transverse'' coordinates  that determine 
the semiclassical string profile  and also 
the structure 
of the coherent operator  on the  SYM side, 
$\tr( \Prod_\s  u_i(\s)  \Phi_i)  $ (see \ci{kru,kt}  and below). 

The agreement between the low-energy effective actions
 on the spin 
chain and the string side explains not only 
the matching of  
energies of  coherent states representing 
 configurations with two
large spins (and also the matching of near-by fluctuations) but
 also the  equivalence of the 
  integrable structures  (which was observed on specific 
examples in \ci{as,Min2}).

%%%%%%%%%%%%%%%%%%%%%%%%%%%%%%%%%%%%%%%%%%%%%%%
\subsection{Higher orders in $\l$}
%%%%%%%%%%%%%%%%%%%%%%%%%%%%%%%%%%%%%%%%%%

The above  leading-order agreement in $SU(2)$ sector has several
generalizations. 

First, we may include higher-order terms 
on the string side. Expanding 
\rf{eaa} in $\td \l$ and eliminating higher powers of time
derivatives  by field redefinitions (which can be done since the  
leading-order equation of motion is first order in time derivative)
we end up with \ci{krt} ($n'\equiv \del_\s n$) 
$$
\LL= \vec C \cdot  \dot{  \vec n }- 
{\tl \ov 8}   \vec n'^2  
+ { \tl^2 \ov 32}  ( \vn''^2 - { 3 \ov 4} \vn'^4) $$
\be \la{thr}
- { \tl^3 \ov 64}  \big[ \vn'''^2 - { 7 \ov 4} \vn'^2 \vn''^2 
- {25 \ov 2} (\vn'  \vn'')^2 + { 13 \ov 16} \vn'^6\big] 
+  O(\tl^4)    \ . \ee
The same $\tl^2$ term is obtained \ci{krt} 
in the coherent state 
action on the spin chain side by starting with the 
sum of the 1-loop dilatation operator \rf{ferr} 
 and the   2-loop term found in   \ci{bks}
\be \la{two}
D_2 = { 2 \l^2 \ov (4 \pi)^4} \sum_{l=1}^J ( Q_{l,l+2} - 4
Q_{l,l+1}) \ , \ \ \ \ \ \ \ \ 
Q_{k,l} \equiv \ha (I - \vec\s_k \cdot \vec\s_l)  \ . \ee 
The $\vn'^4$ term originates from a non-trivial 
quantum correction on the spin chain side. 
This explains  the matching of energies and dimensions 
to the first two orders in $\tl$,  observed on specific 
examples (using the generalized 
Bethe ansatz on the spin chain side) 
 in \ci{SS}. 
The equivalent general conclusion about 2-loop matching was 
obtained in the integrability-based approach in \ci{kmmz}.

The order-by-order agreement seems  to break down 
at $\tl^3$ term. A  natural explanation   
 \ci{SS,bds}
 is that the string limit (first $J \to \infty$, then 
 $\tilde \l \to 0$) and the SYM limit 
 (first $\l \to 0$, then  $J \to \infty$) need not
 produce the same result when applied to the 2-parameter functions 
 $E=\Delta(\l, J)$. 
 A proposal  of  how to  ``complete''
 the perturbative 
 gauge-theory expression  to  make  the agreement 
 with string theory manifest appeared in  \ci{bds}; 
 ref. \ci{afs} also suggested  a
 possible form of   generalized Bethe ansatz on the string side 
that would  naturally  interpolate between the string and gauge
theory results (see also \ci{beitt}). 
 These proposals 
 also suggest a resolution of  the order $\tl^3$ 
  disagreement \ci{cal} 
  between the string theory and gauge theory 
  predictions for $1 \ov J$ corrections to the BMN spectrum.
  
  A possible explanation of why we found the  agreement
  of $\tl$ and $\tl^2$ terms  
  is that the structure of the dilatation operator  at one and
  two loop orders  is, in a sense, 
    fixed by the BMN  limit, 
  which thus essentially determines the low energy effective
  action  in a unique way. This is no longer so starting with
  the 3-loop order, where the dilatation operator  
  already contains \ci{bks}  the 4-spin  $Q Q$ interactions 
   (cf. \rf{two}) 
    which do not contribute 
   to anomalous 
    dimensions in the strict BMN limit. 
    
By analogy  with  non-renormalization (due to underlaying
supersymmetry)
of few leading terms in low-energy effective actions,  
one may suggest that here the 3-loop problem may be 
related to the appearance of non-trivial interpolating functions
as coefficients of the $Q^n \ (n \geq 2) $ terms in the 
dilatation operator. This would,  in particular,  explain 
why    different  structures of the $QQ$ terms  are 
found in the  ``gauge theory''  \ci{bks}
and ``string theory'' \ci{beitt} 
 limits.

 As was
suggested in \ci{SS,bds}, the  disagreement 
between the  string and gauge theory results at  3-loop
order in $\l$ and the leading order in $J$ 
can be repaired   by
adding ``wrapping''   contributions to the dilatation operator
(and thus to the Bethe ansatz relations) on the gauge theory side.
To illustrate this possibility, let us consider   the circular solution 
 case \ci{ft2},  and use the function  $
\l^J\ov (1 + \l)^J$ \ci{bds},  
which is equal to 1 in the string theory  limit ($J\to
\infty$ with fixed ${\l\ov J^2} \equiv \tilde \lambda$)  but zero in
the perturbative gauge theory limit, 
in  order  to interpolate between the
different ${\l^3  \ov J^5}$  coefficients as follows:
\be \la{qu}
 \Delta = J + {  \l \ov 2 J}  - { \l^2 \ov 8 J^3} +
   { \l^3  \ov 16  J^5} { \l^{J-3} \ov (1 + \l)^{J-3}}  + ... \ .
\ee
This expression  agrees with both the string result 
($E=\sqrt{J^2 + \l}$ in \rf{eee} with $n=1$)  and the
perturbative gauge theory  result ($\Delta_{\rm pert}=   
J + { \l \ov 2 J}  - { \l^2 \ov 8 
J^3} + 0\times  \l^3 + ...$  \ci{SS}).

%%%%%%%%%%%%%%%%%%%%%%%%%%%%%%%%%%%%%%%%%%%%%%%%%%%%%%%%%%%%%%%
   
Let us add few details  about  the 
coherent-state expectation value
of higher-loop SYM  dilatation operator.   This expectation value 
appears in  the action in 
the coherent state path integral \rf{zz}
of the quantum spin chain.
Written in terms of independent permutations 
or  
$Q_{l,k} \equiv  I - P_{l,k} = \ha ( 1 -   \s_l \cdot \s_k)
$
the ``$r$-loop'' term in the planar  dilatation operator 
is expected \ci{bks,bei} 
to contain $Q$ in maximal power 
 $[{r+1\ov 2}]$, i.e. $D_1, D_2 \sim \sum Q$,\ \  
$D_3,D_4 \sim \sum  Q + \sum QQ$, etc. 
Explicitly, 
\be \la{dod}
 D=
\sum_{r=0}^\infty { \l^r \ov (4 \pi)^{2r}}   D_{r}  \ , \ \ \ \ \
\ \ \ \ \ \ \ \ \ \
 D_{r} =   \sum^{J}_{l=1} \cD_{r}(l) \ ,
\ee
where as in \rf{ferr} \ci{mz1}  and \rf{two} \ci{bks} 
\be \la{ii}
\cD_0 = I\ ,
\ \ \ \ 
 \cD_1= 2 Q_{l,l+1} \ , \ \ \ \
\cD_2=  - 2( 4 Q_{l,l+1}  -Q_{l,l+2})   \ . \ee
The 3-loop term \ci{bks,bei}
is a special case  of a 2-parameter 
family  \ci{beit,beitt}
$$
 \cD_3(\a,c)  =4(15Q_{l,l+1}    -6  Q_{l,l+2}  +  Q_{l,l+3 })
 $$
\be \la{yity}
+ \   b_1   Q_{l,l+2} Q_{l+1,l+3}  
 + b_2  Q_{l,l+3} Q_{l+1,l+2} +  b_3  Q_{l,l+1} Q_{l+2,l+3}     
   \ ,  \ee
\be \la{nume}
b_1 = 4 + 2 c - \a \ , \ \ \ \ \ 
b_2 = -4 - 4 c +  \a \ , \ \ \ \ \ 
b_3 =  - 2 c  + 5 \a \ .  \ee 
The choice  of $c=0, \ \a=0$ corresponds 
to the  3-loop gauge theory  operator
 of \ci{bks}  whose form  is fixed 
   \ci{bei} by the superconformal
 algebra  and the structure of Feynman diagrams
 (and the BMN scaling). 
 This   choice is also consistent with integrability of the spin chain.  
 To preserve integrability \ci{beit} one should set $\a=0$ (this is the
parameter $\a_2$ of \ci{beit}), 
while to have  consistency  with the gauge-theory
perturbative expansion \ci{beitt}
 one should set $c=0$ ($c\equiv c_4$ in 
\ci{beitt}). 
The case of $\a=2, \ c=0$ corresponds to the  operator mentioned 
in \ci{beit}
which seemed to agree with some string-theory results. 
The case of $\a=0, \ c=1$ is the ``string'' operator \ci{beitt}
which should correspond to the ``string'' modification of the 
Bethe ansatz equations in \ci{afs}.

Starting with an $SU(2)$  coherent state satisfying \rf{coha}  
for which 
\be  \la{rty}
\bra{n}Q_{l,k} \ket{n} = \ha (  1 - n_l \cdot n_k) 
  = \four  ( n_l - n_k)^2 \ , 
 \ee
computing $ \bra{n} D \ket{n}$ and then 
   taking  the   continuum limit
(by introducing a 
spatial coordinate $0< \s\leq 2\pi$, and
 $n(\s_l)= n({ 2\pi l \ov J})$, so that
 $n_{l+1} - n_{l} =  { 2\pi\ov  J} \d_\s n + \ldots$, etc.)
we find, using Taylor expansion and
dropping a total derivative over $\s$, see \ci{krt}, 
\be \la{lety}
{ \l \ov (4 \pi)^2} \bra{n}\cD_1\ket{n} \  \
 \to  \ \  { \bl \ov 8}   \bigg[ \vn'^2 + O({1 \ov J^2} \del^4 \vn)
 \bigg] \ , \ \ \ \ \ \ \ \ \
\  \bl = {\l\ov J^2} \ ,  \ee
\be \la{wwty}
{ \l^2 \ov (4 \pi)^4}  \bra{n}\cD_2\ket{n}\ \ \to \ \
 - { \bl^2 \ov 32}     \bigg[  \vn''^2 + O({1 \ov J^2} \del^6 \vn)
 \bigg]  \
 \ , \ee
 $$
{ \l^3 \ov (4 \pi)^6} \bra{n}\cD_3\ket{n}
  \  \to \ \  { \bl^3 \ov   64}\bigg[\  k_1 { J^2 \ov (2\pi)^2 } 
   \vn'^4
$$ \be\la{iqty}    +   \   \vn'''^2
+ \ k_2  \vn'^2 \vn''^2 
+  \ k_3   (\vn'  \vn'')^2
 + O({ 1 \ov J^2 }\del^8 \vn )\bigg] \ .  \ee
 Here 
 $$
 k_1 = { 1 \ov 16} ( 16 b_1 + 9 b_2 + b_3) =  { 1 \ov 8}
   ( 14  - 3 c  - \a)
 $$
 $$ 
  k_2 = - { 1 \ov 96} (  64  b_1 + 45 b_2 + b_3) = - { 1 \ov 48} 
   ( 38   - 27  c  - 7  \a) \  , $$
    \be \la{nuu} 
     k_3 = - { 5 \ov 48} (  32 b_1 + 9 b_2 + 5 b_3)=
     -  { 5 \ov 24}  ( 46   + 9 c   + \a) \  .\ee
 Note a  relation:
 $   k_1 +   k_2  + { 1 \ov 10}  k_3=0$. 
  The  problematic
 scaling-violating term  $J^2 (\del^2_1 n)^4$ in \rf{iqty}
 does not cancel automatically; it should  cancel after 
 one takes into account quantum corrections (which survive 
 the continuum limit beyond the order $\tl$ approximation \ci{krt}). 
 Quantum corrections are expected also to be  important 
 in order to demonstrate the equivalence of the spin-chain result 
 (for the ``string'' choice of $c=1, \ \a=0$) 
 with the string-theory result in \rf{thr}. Verifying  this
 equivalence  beyond the quadratic $\vn^2$-terms (which obviously agree) 
 remains an open problem. 
  
 Let us also mention  that one can sum up all terms in the 
 string effective
 Hamiltonian  that are of second order  in $\vn$
 but to arbitrary order in $\s$-derivatives \ci{krt} 
 \be \la{gy}
\LL= \vec C \cdot  \dot{  \vec n } - { 1 \ov 4} \vn \left( \sqrt{1 - \tl\ 
\del^2} -1\right)  \vn  +  O(\vn^3) \ . \ee
This expression is in agreement with the leading-order results \rf{thr}
and with the exact BMN  spectrum (see also
 \ci{rt}). The  coherent analogs of BMN states  correspond to 
 small fluctuations near the vacuum state $\vn_0 = (0,0,1)$. 
 On the spin chain side these correspond (in discrete version) 
 to the microscopic spin wave excitations or magnons.

  %%%%%%%%%%%%%%%%%%%%%%%%%%%%%%%%%%%%
  \subsection{Other sectors}
  %%%%%%%%%%%%%%%%%%%%%%%%%%%%%%%%%%%%%%%%%%%%%%%%%%%%%%%%%%%5
  
 It is possible  to extend 
 the approach of~\ci{kru,krt}
to other sectors of rotating string states \ci{lopez,ST,kt}.
 First, one is to  identify  subsectors of operators of 
  the SYM  theory which are closed under renormalization 
  at least to  one-loop.  
  Such   bosonic subsectors 
    are:

(i) the three-spin 
``$SU(3)$'' sector 
 of string configurations  with  all 
three $S^5$ angular momenta 
$(J_1,J_2,J_3)$ being non-zero. They are 
dual to  chiral operators 
$\tr (\Phi_1^{J_1} \Phi_2^{J_2} \Phi_3^{J_3}) + ... $.
These  form a set closed only under one-loop renormalization~\ci{bks}, 
but in the limit when $L= J_1+ J_2 + J_3$ is 
large one can treat this sector as closed even beyond  one loop 
(mixings with fermionic operators  are suppressed by $1/L$
\ci{Min3}).

(ii) the ``$SO(6)$'' sector of  generic (pulsating and rotating)
string motions in $S^5$  which are dual to 
operators built out of 6 real  scalars.  
Examples of  pulsating string states were considered in 
  \ci{Min1,Min2}  and  more generally in \ci{kt};
  this sector  will be discussed 
  in the next section. 
 As was  pointed out  in \ci{Min3}, one can also 
 consider $SO(3)$ and $SO(4)$ subsectors  of the $SO(6)$ sector 
 which are again closed modulo $1/L$ corrections.

(iii) the two-spin  ``$SL(2)$'' sector of string
configurations  with one $AdS_5$ spin $S=S_1$ and one $S^5$ 
angular momentum $J=J_3$, which are dual to operators 
$\tr (D_{1+i2}^S \Phi^J) +...$ (forming a set  closed 
under renormalization to all orders~\ci{BS,kog}).

(iv) the three-spin ``$SU(1,2)$'' sector of string
 configurations with two $AdS_5$ angular momenta 
 $S_1,S_2$ and one $S^5$ 
spin $J=J_3$,  dual to operators 
$\tr (D_{1+i2}^{S_1}D_{3+i4}^{S_2} \Phi^J) +...$ 
which form  a set closed under one-loop renormalization.

 Operators  carrying more general
 combinations of non-zero spins from the list 
  $(S_1,S_2,J_1,J_2,J_3)$ 
 mix with fermionic  and field-strenth 
 operators already  at one loop 
 and would require to consider the full 
 superspin  chains \ci{BS,bb};
  on the string side one would then 
 need to  include fermions of the GS action \ci{mih3}.
 It may happen that  one can again isolate
 some more general  subsectors closed modulo $1/L$ corrections, 
 but it appears that in order to apply a derivation of the 
 reduced sigma model action on the string side one 
 would need to impose additional constraints  on the 
 form of string  configurations \ci{ST}.

 It is indeed straightforward  to  generalize \ci{lopez,ST} the 
  leading-order agreement  observed in the $SU(2)$ sector to
 the $SU(3)$ sector of states with three large $S^5$ spins
 $J_i, 
 \ i=1,2,3$,  finding the  $CP^2$ analog of the 
 $CP^1$   Landau-Lifshitz
  Lagrangian in \rf{con},\rf{coon} 
   \ci{ST} 
  $\LL = - i u^*_i \del_0 u_i - \ha \tl |D_1 u_i|^2 $
  on both the string and the spin chain sides. 
  
  Similar conclusion is reached  \ci{ST} (see also \ci{koz}) 
  in the $SL(2)$ sector of
  $(S,J)$  states  dual to operators like 
  $\Tr (D^S \Phi^J) + ...$. Like in the $SU(2)$ sector
  here the one-loop 
   dilatation operator $D_1$   may be interpreted as the XXX$_{-1/2}$
   Heisenberg spin chain Hamiltonian   \ci{BS}.
   The corresponding coherent states 
   (related to the $SU(2)$ ones by
   an ``analytic continuation'' from the 2-sphere to 
   the 2-hyperboloid,  
   $\vn \to \vec \el,
    \ \eta^{ij} \el_i \el_j = 
  -\el^2_1 + \el^2_2 + \el^2_3 =-1$)
    are defined so that for the $SL(2)$
   generators $S_i$ 
    one has $\bra{\el} S_i \ket{\el} = - \ha  \el_i $
   and then \ci{ST}
 $$
\bra{\el}  (D_1)_{_{SL(2)}}
\ket{\el}= {2 \l \ov ( 4 \pi)^2} 
\sum_{l=1}^J 
\ln  [ \ha  ({ 1 - \eta_{ij} \el^i_{l} \el^j_{l+1} }) ]    $$
\be \la{imp}
= {2 \l \ov ( 4 \pi)^2} 
\sum_{l=1}^J 
\ \ln \big[ 1 + { 1 \ov 4}  \eta_{ij}( \el^i_{l} - \el^i_{l+1})
( \el^j_{l} - \el^j_{l+1}) \big] 
  \ . \ee
 Note that this   is the direct $(-++)$ signature 
  analog on the { classically}
   integrable  lattice Hamiltonian for the Heisenberg magnetic
   \ci{fad}. 
 Since we are interested in comparing to the 
semiclassical  string  case,  $S$ 
as well as $J$  should be large, 
and,  in view of the  ferromagnetic nature of the spin chain, 
 this effectively amounts   to a low-energy  semiclassical limit of 
the chain.
 Considering  the  limit 
$J\to \infty$ with fixed $\tl = { \l\ov J^2}$
 we get, as in the $SU(2)$ case, 
$\el(\sigma_l)= \el(\frac{2\pi l}{J})\equiv \el_l $
and $   \el_{l+1}-\el_l=
\frac{2\pi}{J}\del_\s  \el +{\cal O}(\frac{1}{J^2} \del^2_\s \el).
$
To the one loop order, i.e. with  only one power of $\l$ in \rf{imp}, 
   in expanding the logarithm we need to keep only 
   the order $1\ov J^2$ term, 
   i.e. the term quadratic in first derivatives. 
 This leads to 
 \be \la{kou}
\bra{\el}(D_1)_{_{SL(2)}}\ket{\el} \to 
J\int_0^{2\pi}\frac{d\sigma}{2\pi}\ \bigg[\  \frac{{\tilde\lambda}}{8}
 \eta_{ij}  \el'^i \el'^j +
   {\cal O}(\frac{1}{J^2} \del^4 \el) \bigg] \,.
\ee
As in the $SU(2)$ case, the 
 same term in the action (containing also   the WZ term)
is found  on the string theory side.
 This  
 implies   the general agreement between the string 
 and SYM theory predictions for the energies/dimensions
 at leading order in $\tl$ in the $SL(2)$ sector and thus 
 generalizes the  previous results \ci{bfst}
 found  for  particular solutions.

%%%%%%%%%%%%%%%%%%%%%%%%%%%%%%%%%%%%%%%%%%%%
\setcounter{equation}{0}
%%%%%%%%%%%%%%%%%%%%%%%%%%%%%%%%%%%%%%%%%%%%
%\setcounter{footnote}{0}
\renewcommand{\theequation}{\thesection\arabic{equation}}
%\renewcommand{\thefootnote}{\fnsymbol{footnote}}

%%%%%%%%%%%%%%%%%%%%%%%%%%%%%%%%%%%%%%%%%%%%%%%%%%%%%%%%%%
\section{General fast motion in $S^5$ \\ and 
$SO(6)$ scalar operators}
%%%%%%%%%%%%%%%%%%%%%%%%%%%%%%%%%%%%%%%%%%%%%%%%%%%%%%%%%%%%%%%

One would like to try to understand the general conditions on
string states  and SYM operators  for which the above 
correspondence works, and  incorporate also 
states with large oscillation numbers.
Here we will follow \ci{ST,kt} (a closely related approach was
developed in \ci{mik,mikk,mih3}).
For strings moving in $S^5$ with large oscillation  number 
the energy is 
$E= L + c_1 { \l \ov L} + ...$, i.e.  it is again regular 
in the limit 
$L \to \infty, \ \tl= { \l \ov L^2}\to 0$ 
\ci{Min1}. The  
leading-order duality 
relation between string energies  and 
anomalous dimensions in this case 
was  checked in \ci{bmsz,Min2,Min3}.
The general condition on string solutions  for which 
$E/L= f( \tl)$ has a regular expansion in $\tl$ 
appears to be that the world sheet metric should 
degenerate \ci{mik} in
the 
$\tl \to 0$ limit, i.e. the string motion should be  
ultra-relativistic 
in the small effective string tension limit 
(in the strict tensionless limit  the corresponding states
become BPS \ci{mateos}). 

For example, in the conformal gauge 
 the  2-d induced metric in general  scales as 
$g_{00} = - \kappa^2 + ...$ (assuming  $t=\kappa \tau$, etc.), 
 or, after a rescaling of the 2-d time 
coordinate,  $g_{00} = -1 + O(\tl) + ...$, where  we used that 
$\kappa = {1 \ov \sqrt{\tl}}$. For the fast-moving strings, the leading 
$O(1)$ term in the metric gets cancelled out, and thus the metric degenerates 
in the $\tl \to 0$ limit.

In the strict tensionless 
$\tl\to 0$ limit each string piece is following a
geodesic (big circle) in  $S^5$, while switching on tension
leads to a  slight deviation from geodesic flow, i.e. to a 
nearly-null world surface \ci{mik}.  
The dual coherent SYM operators are then ``locally BPS'', 
i.e. each string bit corresponds to a BPS linear combination 
of 6 scalars (see below).
 In general, the  scalar operators 
can be written as \be \O=  C_{m_1 ... m_L} \tr ( \p_{m_1} 
... \p_{m_L}) \ . \ee 
The planar 1-loop  dilatation operator $D_1$ 
acting on $ C_{m_1 ... m_L}$
 was found in \ci{mz1} to be   equivalent 
to an integrable $SO(6)$ spin chain Hamiltonian
\be
\rH_{m_1\cdots m_L}^{ n_1\ldots n_L} = \frac{\lambda}{(4\pi)^2} 
  \sum_{l=1}^{L} \left(\delta_{m_lm_{l+1}}\delta^{n_ln_{l+1}}+
4\delta^{n_l}_{[m_l}\delta^{n_{l+1}}_{m_{l+1}]}\right)\ .
\label{Hamil}
\ee
To find the analog of the coherent-state action \rf{coon} 
 we choose a natural set of  coherent states 
 $\Pi_l \ket{v_l}$, where at each  site 
 $ \ket{v}= R(v) \ket{0}$. Here 
 $R$ is an $G=SO(6)$ rotation 
 and $\ket{0}$ is the BPS vacuum
 state corresponding to $\tr ( \p_1 + i \p_2)^L$ or 
 $v_{(0)} = (1,i,0,0,0,0)$, which is invariant 
 under the  subgroup 
  $H= SO(2) \times SO(4)$. Then the rotation 
 $R(v)$  and thus the coherent state is parametrized by a point in   
 $$G/H=SO(6)/[SO(4)\times SO(2)] \ , $$ 
  i.e. $v$
  belongs to the
 Grassmanian $G_{2,6}$ \ci{ST}.
$G_{2,6}$ is  thus  the coherent state 
target space for the spin chain sigma model since it 
parametrizes the orbits of the  half-BPS operator
$\p_1+i\p_2$ under the $SO(6)$ rotations. 
 This is the space of 2-planes passing through zero 
  in $R^6$, or
  the space of big circles in $S^5$, 
 i.e. the moduli space of geodesics in $S^5$ 
 \ci{mikk}.  It can be represented also as an 8-dimensional
 quadric  in $CP^5$: a complex 6-vector $v_m$ 
 should be subject, in addition to $v_m v^*_m=1$
 (and gauging away the common phase) 
 also to  $v_m v_m=0$ condition. 
 Taking the limit $L\to \infty$ with fixed $\tl={\l \ov 
 L^2} $   and  the continuum limit $v_{lm}(t) \to 
 v_m(t,\s)$ we then get  the $G_{2,6}$  
 analog of the $CP^1$ action \rf{con},\rf{coon} 
 \be \la{ggg}
 \rS= L \int dt \inti \left( 
- i  v^*_m \del_t v_m  -  \ha \tl | D_\s v_m|^2 \right) \ ,\ee
 $$  v_m v^*_m=1\ , \ \ \ \ v_m v_m=0 \ ,  \ \ \ \ \ \ 
 D_\s v_m = \del_\s v_m - (v^* \del_\s v) v_m \ . $$
 This is also a generalization of the $CP^2$ action found in the
 $SU(3)$ sector  \ci{ST}.
 
   One may wonder  how this 8-dimensional sigma model may  be
  related to the string sigma model on $R \times S^5$ where the
  coordinate space  of transverse motions is only 4-dimensional. 
  The crucial point is that the coherent state action is defined 
  on the phase space (cf. the harmonic oscillator case in sect. 2.1), 
  and $8=(1+5) \times 2 - 2 \times 2 $ is indeed the phase space 
  dimension of a string moving in $R_t \times S^5$.
  
  On the string side, the need to use the phase space description is
  related to the fact that to isolate a ``fast'' coordinate $\a$ 
  for a generic string motion we need to specify both the position
  and velocity of each string piece. Given 
  $\LL= - (\del t)^2 + (\del X_m)^2 $ in conformal gauge 
  ($ \dot X X'=0, \ \dot X^2 + X'^2 =\k^2$, \ $X^2_m=1$) 
   we find that the  point-like trajectories
  (geodesics) are described by 
    $$X_m = a_m \cos \a + b_m \sin \a\ , 
  \ \ \ \   \a =\k \tau , \ \ a^2_m=1 , \ \  b^2_m=1 , \ \ a_m b_m=0\
  . 
  $$
  Equivalently, 
  $$X_m = { 1 \ov \sqrt 2} ( e^{i\a } v_m + e^{-i\a } v^*_m)\ , 
   \ \ \ \ \   v_m = { 1 \ov \sqrt 2}(a_m - i b_m), \ 
 \   |v|^2=1,\  \ v^2=0\ , $$
  where  the constant vector $v_m$ thus 
   belongs to $G_{2,6}$. 
   In general, for near-relativistic
   string  motions 
    $v_m$ should  change slowly with $\tau$ and $\s$. 
   Then starting with the phase space Lagrangian for $(X_m, p_m)$ 
  \be \LL= p_m \dot X_m - \ha p_m p_m - \ha X'_m X'_m \ ,  \ee 
   we may  change the variables according to \ci{kt}
   (cf. again the harmonic oscillator case)
\be X_m = { 1 \ov \sqrt 2} ( e^{i\a } v_m + e^{-i\a } 
v^*_m)\ , \ \ \ \ \ 
\ \ \  
   \ p_m = { i \ov \sqrt 2} p_\a ( e^{i\a } v_m
    - e^{-i\a } v^*_m)\ , \ee
   where $\a$ and $v_m$  now depend on $\tau$ and $\s$
   and $v_m$  again  belongs to $G_{2,6}$. There is an
    obvious $U(1)$ gauge invariance, $\a\to \a-\b, \ v_m
\to e^{i\b} v_m$. Gauge-fixing the 2-d reparametrizations 
by $t \sim \tau, \ p_\a \sim L$
(or, doing an  approximate  T-duality  $\a\to \tilde \a$  
 and  setting 
 $\td \a \sim \s$ as in sect. 2.3)
one finds,  after an additional   rescaling of the time coordinate, 
that the phase-space Lagrangian  becomes  \ci{kt}:
\be 
\LL=  p_\a D_t \a - \ha \tl |D_\s v|^2 - {1 \ov 4}\tl  \big[ e^{2 i \a} 
( D_\s v)^2 + c.c.\big]\ . \ee
 The first term here produces the WZ term 
 $ -i v^*_m \del_t
v_m$  and the last one  averages to zero  since 
$\a \approx \k \tau + ...$,  where $\k= ({ \sqrt {\td \l}  })^{-1} 
 \to \infty$. 
 
Equivalently, the $\a$-dependent terms in the action 
(that were absent in the   $SU(2)$ or $SU(3)$ 
 sectors) 
can be eliminated by canonical transformations \ci{kt}.
We then end up with the following 8-dimensional 
phase-space Lagrangian 
for the ``transverse'' string motions, 
$ \LL = -i v^*_m \del_t v_m  - \ha \tl | D_\s v_m|^2,$
which is the same as found on the spin chain side \rf{ggg}. 
The  3-spin $SU(3)$  case is the special 
case when $v_m=(u_1,i u_1, u_2, i u_2, u_3, i u_3)$, 
where $u_i$ belongs to the $CP^2$ subspace of $G_{2,6}$. 
The agreement between the spin chain and the string sides in this
general $G_{2,6}=
SO(6)/[SO(4)\times SO(2)]$ case explains not only 
the matching for pulsating solutions \ci{Min1,Min2} but also
 for near-by fluctuations.  
 
 %%%%%%%%%%%%%%%%%%%%%%%%%%%%%%%%%%%%%%%%%%%%%%%%%%%%%%
 
 Let us now 
  discuss the reason for the restriction $v^2=0$ on the spin chain
  side 
 and also clarify  the structure of the coherent operators 
  corresponding to
 semiclassical string states. 
 Given the scalar  operator  $\O= C_{m_1\ldots m_L} \tr 
 \left(\p_{m_1}\ldots \p_{m_L} \right) $ one   
 may obtain the Schr\"odinger  equation for 
 the wave function $C_{m_1\ldots m_L}(t)$ from\foot{For the coherent states we consider 
 the corresponding equation of motion 
 may be interpreted as a (non-trivial) 
   RG equation 
for the  coupling constant associated to  the operator 
$ \O$.}
\be 
\rS = -\int dt \bigg(  i 
C^*_{m_1\ldots m_L} \del_t  C_{m_1\ldots m_L} 
  +
C^*_{m_1\ldots m_L} \rH_{m_1\cdots m_L}^{
 n_1\ldots n_L} C_{n_1\ldots n_L} \bigg) \ . \ee 
 In the limit $L\to \infty$  we may  consider
 the coherent state description  and assume the  factorized ansatz 
 \ci{kt}
 \be C_{m_1\ldots m_L}= v_{m_1} ... v_{m_L}  \ , \ee
  where each 
 $ v_{l}= \{v_{m_l}\}$ ($l=1,...,L$) 
 is a complex unit-norm 6-vector.
 The BPS case corresponding to the totally symmetric traceless 
 $C_{m_1\ldots m_L}$ is represented by $ v_{l}=  v_{ (0)}$, 
 $v_{(0)}^2 =0$.  Using \rf{Hamil} and substituting the 
 ansatz for $C_{m_1\ldots m_L}$ into the above action one finds 
 \be \la{acta}
\rS = - \int dt \sum_{l=1}^{L} \bigg( i \bv_l \del_t  v_l + 
   { \l \ov (4\pi)^2} 
\bigg[(\bv_l\bv_{l+1})(v_lv_{l+1}) + 2 
 - 2 (\bv_lv_{l+1})(v_l\bv_{l+1})\bigg]
 \bigg)  , \ee
 where we suppressed the 6-vector indices in the scalar products.
As expected \ci{mz1}, the coherent state expectation value of the  
Hamiltonian (i.e. order $\l$  term in \rf{acta}) 
 vanishes  for the 
BPS  case when $v_l$ does not depend on $l$ and  $v^2=0$.
More generally, if we assume that $v_l$ is changing slowly
 with $l$ (\ie\ $v_l\simeq v_{l+1}$), 
then we find that \rf{acta} contains a  potential term 
$(\bv_l\bv_{l})(v_lv_{l})$ coming from the first ``trace'' structure in \rf{Hamil}.
This  term will lead  to large (order $\l L$ \ci{mz1}) 
shifts of anomalous dimensions, 
invalidating  low-energy  expansion, i.e. prohibiting one from taking the  
continuum limit  $
L\to \infty , \   \tl={\l\ov L^2}={\rm fixed}$, 
 and thus from establishing  direct 
  correspondence with string theory 
along the lines of \ci{kru,krt,ST}.\foot{Even $SU(2)$ sector
 has in general 
  other higher-energy  states with dimensions 
  $ \sim \l L$  (in addition to magnons with energies 
  $\sim { \l  \ov L^2}$    and macroscopic spin waves 
  with $\Delta \sim { \l  \ov L}$
  there are also spinons with $\Delta \sim { \l  L}$)
  but these do not correspond to fast strings -- they  are 
  not captured  by  
  the low-energy continuum limit of the coherent state 
  path integral.}
  
To 
get solutions
with  small variations of $v_l$ from site to site  we are thus 
to impose \be 
v^2_l=0\  , \ \ \ \ \ \ \ \ \ 
l=1,\ldots,L \ee
 which minimizes the potential energy coming 
from the first term in  (\ref{Hamil}).
This  condition 
implies that the operator at { each} site is 
invariant under half of supersymmetries:
if $v^2=0$ the matrix $v_m\Gamma^m$ appearing in the 
 variation of the 
operator  $v_m \p_m $, i.e. $
 \delta_\epsilon (v_m  \p_m )= \frac{i}{2} 
 \bar{\epsilon} (v_m \Gamma^m) \psi $, 
 satisfies $(v_m\Gamma^m)^2=0$.
This means    that 
$v_m \p_m $ is invariant under the  variations 
associated with the null eigenvalues \ci{kt}.
One  may thus call $v^2=0$   a ``{\it locally BPS}'' condition
 since the preserved combinations of supercharges 
in general are different for each $v_l$, i.e.  the 
operator corresponding to  $C=v_1... v_L$ is not BPS.
Here ``local'' should be
 understood in the sense of the spin chain,  or,  
equivalently,  the spatial world-sheet direction.\foot{This 
generalizes the argument implicit in \ci{kru}; an 
equivalent proposal was made in \ci{mikk}. This is related to but different 
from the ``nearly BPS'' operators  discussed in \ci{mateos} 
(which,  by definition,   were those which become BPS in
 the limit $\l\to 0$).}
 
In the case when 
 $v_l$ are slowly changing   we
can take the continuum limit   as in \ci{kru,krt,ST}
by  introducing the 2-d field $v_m(t,\s)$ with 
$v_{ml} (t)= v_m(t, {2\pi l\ov L})$. Then  
 \rf{acta} reduces to \rf{ggg}
(all higher derivative terms are suppressed 
by powers of $1\ov L$ and the potential term 
is absent due to the condition $v^2=0$). Then  
\rf{acta} becomes 
  equivalent to the $G_{2,6}$  Landau-Lifshitz
 sigma model \rf{ggg}
 which was derived from the phase space action  
 on the string side.
 %\foot{The presence of the trace in the SYM operators 
% implies that 
%we have to consider only spin chain states that are 
%invariant under translations in $l$ or in $\sigma$.
% This  means that the momentum in the 
%direction $\sigma$ should be quantized, 
% $P_\sigma=2 \pi n $, or, equivalently, 
%$ \int_0^{2\pi} \frac{d\sigma}{2\pi} 
%\ v^*_m  \partial_\sigma v_m 
% =2 \pi n $. 
%This should be viewed as  a condition on the 
%solutions $v_m(t,\s)$. The same condition appears 
%on the string side from conformal gauge  constraint.}

To  summarize, considering ultra-relativistic strings in $S^5$ one
can isolate a fast variable $\a$ (a ``polar angle'' 
in the string phase space) whose momentum $p_\a$ is large.
One may gauge-fix $p_\a$ to be constant $\sim L$ or set 
 $\td \a \sim L
\s$,  so that $\s$ or the ``operator direction'' 
on the SYM side 
gets interpretation  of  ``T-dual to fast coordinate'' direction. 
As a result, one finds 
 a local phase-space action with 8-dimensional 
target space (where one can not  eliminate 4 momenta 
without spoiling the locality). This action 
is equivalent to the Grassmanian $G_{2,6}$ 
Landau-Lifshitz  sigma model 
action appearing on the 
spin chain side.

As a by-product, 
we  thus get a precise mapping between 
 string solutions and  SYM 
 operators representing coherent spin chain states \ci{kru,kt}. 
 Explicit examples corresponding to pulsating 
and rotating solutions were given in \ci{kt}. 
In the continuum limit we may write  the operator 
 corresponding to the solution $v(t,\s)$ as 
$\O =  \tr [ \prod_\s  \vv(t,\s) ]   ,\  
\vv \equiv  v_{m}(t,\s)  \p_{m}.$
This locally BPS coherent operator is the 
  SYM operator naturally associated to a
ultra-relativistic string solution.
The $t$-dependence of the string solution thus translates 
into the RG scale 
dependence of $\O$, while the $\s$-dependence describes
 the ordering of scalar field  factors under 
the trace. 

 In general,  semiclassical string states 
represented by classical string solutions 
should be dual to coherent spin chain states or
 coherent operators, 
which are different from the exact eigenstates 
of the dilatation operator 
but which should lead to the same energy or anomalous
 dimension  expressions. 
At the same time, the Bethe ansatz approach \ci{mz1,bmsz,bfst,kmmz}
is  determining the  exact eigenvalues 
of the dilatation operator.
The reason why the two approaches happen to be in agreement 
is that in the limit   we consider  the problem is 
essentially semiclassical, and because of  the integrability 
of the spin chain, its exact eigenvalues are not just 
well-approximated by the classical solutions
but are  actually exactly reproduced by them for $L\to \infty$,
 i.e.  
(just as in the harmonic oscillator or 
 flat space string theory case, cf. also \ci{jev})
the semiclassical coherent state 
 sigma model approach happens 
to be exact.

%%%%%%%%%%%%%%%%%%%%%%%%%%%%%%%%%%%%%%%%%%%%%%%%%%%%%%%%%%
\section{Concluding remarks}
%%%%%%%%%%%%%%%%%%%%%%%%%%%%%%

As discussed above, 
there exists a remarkable  generalization  of
the  near-BPS (BMN) limit to  non-BPS  but ``locally-BPS'' 
sector of string/SYM states (for related reviews see 
\ci{tse1,tse2,beise}).
  It remains to understand better 
 when and why  this direct relation works or fails,  
but  the hope is to use  it as a guide  towards  
 finding    the  string/SYM spectrum 
 exactly in $\l$, at least in a subsector of states. 
 The relation between phase-space action
 for ``slow'' variables  on the string side and the 
 coherent-state action on the SYM (spin chain) side 
 gives a very explicit picture of how string action 
 ``emerges''  on  the conformal 
 gauge theory side (with the central role 
 played by the  dilatation operator).
 This  implies not only an   equivalence between string energies 
 and SYM dimensions (established  to first two orders 
 in expansion in the effective coupling $\tl$) 
 but also a direct relation between the string profiles and the
 structure of coherent SYM operators.

One may try  also  to   use the duality 
 as a tool  to uncover 
 the structure of planar SYM theory to all orders  in 
$\l$ by assuming  the exact correspondence between   particular 
SYM  and string states. 
For example, demanding the consistency with the BMN 
scaling limit (along with  superconformal algebra)  determines 
the structure of the full  3-loop SYM dilatation 
operator in the $SU(2)$ sector \ci{bks,beit}.
One can also use the BMN limit to fix only  
 part of the dilatation 
operator but to all orders in $\l$ \ci{rt}.
Generalizing \rf{ferr},\rf{two}  and the 3- and 4-loop 
expressions in \ci{bks,beit} one  can organize \ci{SS,krt,rt}
the dilatation operator as an expansion in powers of 
$Q_{k,l}= \ha (I - \vec \s_k \cdot \vec \s_l)$ which 
reflect interactions between spin chain sites, 
$$D = \sum Q  + \sum QQ + \sum QQQ + ... \ . $$ 
Here the products $Q...Q$ are ``irreducible'',
i.e.  index of each site  appears only once. The $Q^2$-terms first
appear at 3 loops, $Q^3$-terms -- at 5 loops, etc. 
\ci{bks,beit}. 
Concentrating on the order-$Q$ part $D^{(1)}$
 of $D$ one can write:
  \be \la{geg}
  D^{(1)}=\sum_{r=0}^\infty { \l^r \ov (4 \pi)^{r}}  
  \sum^{L}_{l=1} \De_{r}(l)\ , \ \ \ \ \ \ \ 
 \De_{r}(l) = 2\sum_{k=1}^r a_{r,k} Q_{l,l+k}\ ,   \ee
 or 
 $
 D^{(1)}=   \sum_{l=1}^L
 \sum_{k=1}^{L-1} \ h_k (L,\l)\  Q_{l,l+k} .$
 Demanding the agreement with the BMN limit 
  one can then determine the
 coefficients $a_{r,k}$ and thus the function 
  $h_k$ explicitly to all orders in $\l$ \ci{rt}. 
  In particular, for  large $L$, i.e. 
  when $D$ acts on  ``long'' operators, one finds 
  \be
\label{D}
D^{(1)} = 2 \sum_{l=1}^L \sum_{k=1}^\infty
f_k(\lambda)\  Q_{l,l+k} \ , \ \ \ \ \ \ \ \ \ \
f_k(\lambda)= \sum_{r=k}^\infty
{\lambda^r \over (4\pi)^{2r}} \ a_{r,l} \ , 
\ee
where  the coefficients $f_k(\lambda)$
can  be summed up in terms of
the standard Gauss hypergeometric function \ci{rt}
\be
\label{co}
f_k(\lambda) =
 \left( {\lambda \over 4 \pi^2}\right)^k
{\Gamma(k-\ha) \over 4\sqrt \pi \ \Gamma(k+1)}
\;\
{}_2 F_{1} (k-\ha, k+\ha; 2 k + 1; -
 {\lambda \over \pi^2})\ , 
\ee
%Explicitly,  this can be written as 
or, equivalently, 
\be 
f_k(\lambda) = { 1 \ov 4 \pi ( 2 k - 1) }
\left( {\lambda \over  \pi^2}\right)^k
\int^1_0 du \left[ { u(1-u) \ov 1 + {\l \ov \pi^2}
u}\right]^{k-1/2}\ . 
\ee 
$f_k$ goes  rapidly to zero   at large $k$, so we
get  a spin chain with short-range interactions. 

One may hope that imposing additional constraints 
coming from correspondence with other string  solutions
(and using recent insights in  \ci{bds,afs})
may help to determine the structure of the 
dilatation operator further.

The function 
$f_k(\l) $ in \rf{co} smoothly interpolates between the usual 
 perturbative expansion 
at small $\l$  and 
$f_k(\lambda)
\sim \sqrt{\lambda}  $ behaviour 
at large  $\l$. The latter is  the expected  behaviour 
of anomalous dimensions of ``long''  operators dual 
to  ``semiclassical'' string  states.  

Similar interpolating functions should 
appear also in anomalous dimensions of other SYM operators, 
though for ``short'' operators the strong-coupling
 asymptotics of the dimensions should be $\l^{1/4}$. 
Let us consider, for example, the following dimension 4 
supersymmetry descendant of the Konishi 
  operator, $\tr ([\Phi_1,\Phi_2]^2)$
  (which  belongs to the $SU(2)$ sector and should 
  have the same anomalous dimension as the Konishi operator).
   The 
  first few terms in the perturbative  $\l$-expansion of its anomalous
  dimension are known to be 
  \ci{ans,bks,klvo,bei}
 \be \la{oni}
\Delta_{_{\l \ll 1} }
 = 4+  3 \le  -  3 \le^2  + {21 \ov 4} \le^3-{ 705 \ov 64} \le^4 +
 O(\le^5) \ ,\ee 
 $$ 
   \le \equiv {\lambda\over 4\pi^2}   \ .  $$
If one would to ignore all non-linear in $Q$ terms in the dilatation 
operator  \rf{geg}, then the resummed expression 
for the anomalous dimension would be \ci{rt}
\be
\label{koni}
\Delta^{(1)} = 4 + { 3\ov 2} \big( \sqrt{ 1 + 4 \le } -1 \big)
\ ,   \ee
which does not, however, have  the expected  large $\l$ 
asymptotics,  
\be \la{hohi}
\Delta_{_{\l \gg 1} } = 2\l^{1/4}  + ...= 2  \sqrt{ 2 \pi} (\le)^{1/4}
+ ...\ . \ee
 Note that one cannot reproduce such  asymptotics with  
 an interpolating expression for $\Delta$  built out of rational 
 functions of  the  square of the 
 effective  string $\le=T^2$:
   while 
 the expansion of a rational function (like $\sqrt{
 \sqrt{ a + b \le} + d } $)
  at small $\lambda$  would have the same form 
 as   \rf{oni}, the  factors of $\pi$  would not match 
 in the strong coupling limit \rf{hohi}.
 
 The reason for
 the above extra  factor  of $\sqrt \pi$ in
 $ \Delta_{_{\l \gg 1}}$ in  \rf{hohi}
 expressed in terms of $\l_{\rm eff}$
  can be understood  following ref. \ci{gub}.
 The Konishi operator  should correspond to 
  the lowest-level scalar 
  string mode. The masses of the \adss string modes   
 are, in general, non-trivial functions of the 
 string tension
 (the corresponding wave equations 
  receive $\a'$-corrections), 
 but, in the large-tension limit, they should 
 simply be  the same as in flat space,
   i.e. (in units where $R=1$)
 \be m^2 = { 4n  \ov \a'} = 4n \sql =8  n \pi  T \ ,
 \ \ \ \  \  T= {\sql \ov 2 \pi}= \sqrt{\le}  \ . \ee 
 Then the standard $AdS_5$   formula 
 $\D ( \D -4) = m^2$ for the dimension 
 of a scalar field with mass $m$ 
 (again, expected to be  valid in the large tension  limit)
 predicts that for $n=1$ \ 
  \be \la{sto}
  \D-2=  \sqrt{ 4 + m^2} \to  m  + ... =  2 \sqrt{
 2  \pi  \sqrt{\le }} + ...\ . \ee
    It thus appears that instead of being a  rational function
   of $\le$,  the dimension $\D$ of the Konishi operator 
   should  be a transcendental function. 
   In fact, the hypergeometric function     like 
   the one appearing in \rf{co}, i.e. 
   ${}_2 F_{1} (a, b; c; - k \le )$, 
      seems to be a natural candidate:
      one can choose its arguments (and an overall coefficient) 
       so that  to match the powers of $\l$ and $\pi$ 
   in both the strong and the weak coupling limits
   (one example with required strong-coupling asymptotics
    has $a=-1/4, 
   \ b=3/4, \ c=3/2, \ k=4$).
   With  $\D \sim {}_2 F_{1}$  (or $\D$ being 
    a  rational function of 
   ${}_2 F_{1}$)  it is  possible also to 
   satisfy the string-theory requirement that 
   the strong-coupling expansion should be organized 
   as an expansion  in powers of the inverse string tension, i.e.
   $\D = \l^{1/4} ( c_1 + { c_2 \ov \sql} + { c_3 \ov (\sql)^2} +
    ...)
   $ (cf. \rf{sto}).  
    
  At the same time,  for ``long'' operators  with large 
 canonical dimensions (like BMN operators or non-BPS operators 
 discussed  in the previous sections)  
 the interpolating functions  appearing 
 in  $\D$  may be simple  rational 
 functions like square roots:  
  here both the weak (gauge theory) 
 and the strong (string theory) expansions
  are organized  in terms of 
 $\le = {\l \ov 4\pi^2}$ with the leading coefficients
  which do not contain extra
 powers of $\pi$ (see also the discussion below eq. \rf{sma}).

%% optional
\begin{acknowledgments}
I would like to thank  the organizers of 
the Carg\`ese 2004  Summer School
 for the invitation to speak. 
I am  grateful to 
G. Arutyunov, N. Beisert, S. Frolov,
 M. Kruczenski, 
 J. Russo, 
 A. Ryzhov and  B. Stefanski 
for  useful discussions and collaboration
 on the work described above. 
This  work  was supported in part  by the  DOE
grant DE-FG02-91ER40690, the INTAS contract 03-51-6346
and the RS Wolfson award. 
\end{acknowledgments}

%% appendix optional
%\chapappendix{A}

%This is an appendix with a title.
%\chapappendix{}
%This is an appendix without a title.

\end{document}